# Relaxing strong compatibility at atomistic-continuum interface: Introduction to consistent linear coupling method

Pouya Towhidi[*], Manouchehr Salehi

*Department of Mechanical Engineering, Amirkabir University of Technology, Tehran, Iran*

## Abstract

The most essential concept in concurrent multiscale methods involving atomistic-continuum coupling is how to define the relation between atomistic and continuum regions. A well-known coupling method that has been frequently employed in different concurrent multiscale methods such as quasicontinuum is the strong compatibility coupling (SCC). Although the SCC is a highly accurate coupling method, it constrains the mesh generation and restricts the reduction of computational cost. In this paper, first, we explain the notion of coupling models through the interface in the context of continuum mechanics for quasi-static problems. Then, the SCC is relaxed to overcome the downsides in the following approaches: a surface approximation approach and a force approximation approach. Based on the latter, we develop a brand new coupling method called consistent linear coupling (CLC). Overall, six different coupling schemes are introduced in line with the surface and force approximation approaches. The schemes are compared in terms of solving quasi-static 3D elastic nanoscale contact between an aluminum substrate and a diamond semi-sphere. Numerical results reveal that the CLC-based schemes are more accurate when compared with the surface approximation schemes. Furthermore, we demonstrate that by decreasing the size of interface elements, the accuracy of the CLC converges to and even exceeds the accuracy of SCC, but nevertheless with significantly less computational cost. In other words, in contrast to the SCC, the accuracy of the CLC is tunable, and it can be optimized proportional to the computational cost and accuracy. Finally, we sketch the idea of how the CLC can make the consistency near the atomistic-continuum interface by eliminating ''ghost forces.'' The energy of a one-dimensional atomic chain with a finite range of interaction is relaxed to validate the CLC method.

## 1. Introduction

Continuum mechanics is considered a principal model for most engineering applications. The accuracy of the model in comparison to experiments is outstanding, although it is built on the continuity of matter assumption, which is not true due to the discrete nature of matter. The key to success of continuum mechanics is the fact that the dimensions of a medium are many orders of magnitude larger than atomic spacing, which makes the assumption to be valid at the macroscale. With advances in nanotechnology, computational methods are required to predict material behavior accurately. Since, in this case, the continuity assumption in continuum mechanics may no longer be valid, the results of the analysis could be unrealistic, and hence applying atomistic methods instead seems inevitable. For instance, continuum methods may fail in the nanoscale contact when the surface roughness of the tip is defined at the atomic scale [1]. Likewise, investigation of mode I fracture in silicon under tensile or bending loading at the nanoscale led to a specific length scale limit [2], below which breakdown of continuum fracture mechanics occurred. Another example is related to a study of predicting amplitude and wavelength of graphene membrane, rippled by compression (induced by temperature) at nanoscale level [3], in which continuum description resulted in substantial error compared with the experimental data. By contrast, a good agreement was observed between the atomistic tight-binding model and the experiment. The fact that the atomistic models predict phenomena more accurately and fundamentally at the nanoscale is the main reason for the wide applications of molecular dynamics or statics for studying phenomena at this level. Some reviews of these applications on popular topics in solid mechanics are nanoindentation [4], defects [5], and crack propagation [6]. Although atomistic models are considered accurate enough for analysis at nano/microscale, practically speaking, they are prohibitively expensive to apply even at a few micrometers with supercomputers, which seems essential for some engineering applications [7, 8]. In addition, continuum mechanics is acceptable on the whole except in small regions of a domain, such as singularities, defects, and chemical reactions [9]. Unfortunately, these small regions confine the application of continuum mechanics. Limitations of continuum mechanics, in addition to the excessive computational cost of the atomistic models, have caused motivation for the advent and development of concurrent multiscale methods, particularly atomistic-continuum coupling methods, over the past three decades.

---

[*] Corresponding author.
*E-mail address*: p.towhidi@aut.ac.ir (P. Towhidi).

Concurrent atomistic-continuum coupling methods endeavor to take advantage of both models, i.e., the accuracy of the atomistic model and the low computational cost of the continuum model, to provide an efficient computational tool for facing real-world engineering problems at nano/microscale level, e.g., nanoindentation [10, 11], defects [12], crack propagation and fracture [13, 14], MEMS [15], contact [16], and so forth. Thus, they partition a body into atomistic and continuum regions, which makes an interface layer between them. Then, the issue is how to relate these two parts through the interface. One of the earliest approaches for connecting the parts is the strong compatibility coupling (SCC) method, attained by employing a specific mesh generation algorithm and enforcing the displacements of interface nodes with interface atoms. In general, methods that do not use the SCC for connecting atomistic to continuum regions and use other concepts instead are called weak compatibility. Many multiscale methods use the SCC thanks to its accuracy, e.g., the coupled atomistic and discrete dislocation (CADD) method [17, 18], the combined finite element and atomistic (FEAt) [19], the quasicontinuum (QC) method [12, 20], the cluster-based quasicontinuum (CQC) method [21, 22], the coupling of length scales (CLS) method [23]. In contrast, some other methods, such as the bridging domain (BD) method [24] and the atomistic-to-continuum (AtC) method [25, 26], employ a handshake region between atomistic and continuum regions to make a more gradual transition between the two scales [24-27]. Both continuum and atomistic models are present in the handshake region, and the total energy and/or force is calculated as a combination of the energy of both models. In addition, usually, Lagrange multipliers [28] are employed between the nodes and atoms in the handshake region to enforce displacement compatibility. Nevertheless, some other methods use neither the SCC nor the handshake region. Notable examples are the multiresolution molecular mechanics (MMM) method [29], the bridging scale method (BSM) [30, 31], and the composite grid atomistic continuum method (CACM). In MMM, similar to CQC, a nonlocal sampling scheme is performed to compute energy and force in the coarse-grained region as well as in the atomistic region; however, with different assembling energy, and as the authors claimed, this method does not suffer from a significant energy error unlike the CQC method [28]. The BSM is characterized by its unique displacements fields. It is established on the decomposition of displacement on the fine and coarse scales with the orthogonality property between them. Compatibility in the BSM is defined in two ways: (1) The atomistic part affects the continuum part with constraining nodes in the atomistic region by means of the least-squares method. (2) So-called "ghost atoms" are located near the interface in the local side (continuum part) to prepare forces on atoms. According to this compatibility setting, iterative minimization is needed to find a solution [30]. In the CACM, which is noted as a type of Schwarz alternating method [8], atomistic energy functional and continuum energy functional are minimized separately and after each minimization compatibility conditions apply to the other counterpart. Similar to the BSM, this approach requires iterative minimization.

Another important diversity of the methods is based on how coupling methods are formulated, namely force-based or energy-based. The energy-based methods, such as CLS, QC, CQC, BSM, BD, and MMM, concentrate on the total energy of the system and minimize it to find the solution. While the approach seems flawless, natural inconsistency arises near the interface due to the fact that the continuum part is a local model and the atomistic part is nonlocal. This inconsistency exhibits itself as spurious forces or so-called "ghost forces" on the atoms near the interface. The fact that the ghost forces have an adverse effect on the outcome, the force-based methods, such as CAAD, FEAt, and AtC, have gained some attention since it is theoretically possible to eliminate the ghost forces therein. However, they have their own drawbacks. For instance, they do not have conservative energy associated with the forces and hence may converge to unstable equilibrium states.

All of those multiscale methods have their own weaknesses and strengths. Methods with iterative minimization converge relatively late. The cluster-based methods have relatively more computational cost since nonlocal sampling atoms are present throughout a body while they mitigate the ghost forces effect, which could improve the accuracy. Methods without handshake are easier for mesh adaption with regard to converting atomistic region to finite element and vice versa, which seems vital for each multiscale method. On the other hand, mesh generation of the methods with weak compatibility is much easier. The Schwarz alternating type methods, albeit their modularity advantage, converge lately [8] and depend highly on the intersection of two models (atomistic-continuum intersect region) [32]. From the accuracy viewpoint, fourteen multiscale methods were analyzed and compared with each other in the case of the Lomer dislocation dipole [33]. Results showed that methods with strong compatibility are more accurate. This result was confirmed in the case of fracture in nickel [34], although the authors noted that the SCC had more computational cost compared with the weak compatibility methods. The hybrid quasicontinuum method [35] was then proposed to reduce the computational cost of the original version of QC by splitting the continuum region into regions governed by the Cauchy-Born rule and linear elasticity. In particular, the continuum region near the atomistic region was treated by the Cauchy-Born rule, and in the region sufficiently far from the atomistic region, linear elasticity was adopted. In fact, this approach can be applied to the most multiscale methods, and the QC does not surpass the other methods from this perspective. Herein, we are concerned about another source of increasing the computational cost, the strong compatibility itself. Because the size of the elements must be scaled down to the atomic lattice constant at the interface in this method, numerous elements surround the interface, which is related to the computational cost. On the other hand, while early mesh coarsening (starting with large elements) at the interface can significantly reduce the computational cost, it could also reduce the accuracy, which is not desired. Thus, having early mesh coarsening and maintaining the accuracy so far as possible is quite a challenging problem, which we try to address in this paper.

In this study, the main goal is to introduce and discuss some ways to encompass most of the aforementioned advantages. To fulfill this aim, we commence with necessary and sufficient conditions required to be satisfied between inner surfaces of two continuum bodies in order to preserve continuity. Considering the conditions, we relax the strong compatibility within the surface and force approximation approaches. Relaxing the strong compatibility allows early mesh coarsening at the interface. In the surface approximation approach, the continuum surface is approximated by interpolating or curve-fitting methods. In the force approximation approach, the continuum surface is approximated based on the forces on the atomistic surface. We present six schemes based on the surface and force approximation approaches. The common features of the introduced schemes are as follows: no handshake region exists, which makes easier mesh adaption; easy implementation as the mesh generation is not constrained, unlike the SCC; early mesh coarsening is feasible, whereby balancing between the accuracy and the computational cost could be achievable.

We develop a special case of the force approximation approach called consistent linear coupling (CLC). The CLC approximates the continuum inner surface linearly, and it introduces the series of conditions that need to be satisfied in order to produce strong compatibility while retaining early mesh coarsening. For the CLC method, CLC element-based and CLC atom-based schemes are proposed to meet the conditions. The accuracy of the introduced methods is assessed in solving 3D nanoscale contact between diamond indenter and aluminum substrate. Among all introduced methods, CLC-based schemes and the least-squares method are promising due to their superior accuracy. In particular, within the CLC framework, it is possible to reach the same accuracy as the SCC with a considerably lower computational cost.

Ultimately, we discuss another application of the CLC in the atomistic-continuum coupling. We address how it is possible to make a consistency near the atomistic-continuum interface for a one-dimensional atomic chain with an arbitrary range of interaction. As such, we add artificial nodes to the system, located in specified positions and constrained by employing CLC, in order to eliminate ghost forces.

This paper is organized as follows. In Section 2, the basis of our coupling models is presented for a partitioned body in the context of continuum mechanics. In Section 3, the finite element method is adopted for discretizing the continuum body, and also surface and force approximations are described for coupling models. In Section 4, samples of the surface and force approximations are proposed. In the concept of surface approximation, direct coupling, least-squares coupling, and master-slave coupling are formulated. Additionally, the CLC method is developed in an ideal condition based on the force approximation. In Section 5, the problem of the double counting of energy at the interface is investigated, and the total energy of the coupled atomistic-continuum domain is proposed. In Section 6, an elastic nanocontact of a semi-sphere diamond with an aluminum substrate is solved for the proposed coupling schemes, and also a convergence property of the CLC method is investigated by refining mesh near the interface. In Section 7, the CLC method is implemented in a different manner but with the same concept to make consistency in a one-dimensional atomic chain and is compared with a conventional multiscale method. A discussion and conclusions are presented in Section 8 and Section 9, respectively.

## 2. Continuum description for a partitioned body

First, consider a time-independent problem, including a continuous body $B_0$ under prescribed boundary conditions and a body force $\boldsymbol{b}_0$ (Fig. 1). $B_0$ is the body in the reference configuration, and the points therein are denoted by $\boldsymbol{X}$. A static external traction $\boldsymbol{T}_0$ is applied on a part of $B_0$ surface ($\partial B_{0t}$) and $\boldsymbol{\phi}_d$ (essential boundary condition) is determined on the other part ($\partial B_{0d}$), where $\partial B_{0t} \subset \partial B_0, \partial B_{0d} \subset \partial B_0, \partial B_{0t} \cap \partial B_{0d} = \emptyset$ and $\partial B_{0t} \cup \partial B_{0d} = \partial B_0$. Under these conditions, the body undergoes deformation from its initial state. It is essentially a boundary value problem, and the solution to this problem is the deformed configuration of the body.

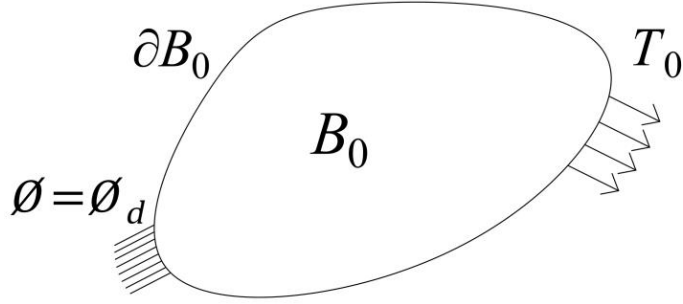

Fig. 1. A typical continuum body in the reference configuration.

Let $\boldsymbol{\phi}$ be the body in the deformed configuration, which is defined as $\boldsymbol{\phi}: B_0 \to \mathbb{R}^3$, and consider a hyperelastic material. The total potential energy functional, $\Pi$, for the described body is written as [36]

$$\Pi(\boldsymbol{\phi}) = \int_{B_0} W(\boldsymbol{F}) \, dX - \int_{B_0} \boldsymbol{\phi}(\boldsymbol{X}) \cdot \rho_0 \boldsymbol{b}_0(\boldsymbol{X}) dX - \int_{\partial B_0} \boldsymbol{T}_0(\boldsymbol{X}) \cdot \boldsymbol{\phi}(\boldsymbol{X}) \, dX, \tag{1}$$

where $\boldsymbol{F}$ is the deformation gradient and is equal to $\partial \boldsymbol{\phi} / \partial \boldsymbol{X}$, $W(\boldsymbol{F})$ is the strain energy density, $\rho_0$ is the mass density in the reference configuration, and recall that $\boldsymbol{b}_0(\boldsymbol{X})$ and $\boldsymbol{T}_0(\boldsymbol{X})$ are the body and traction fields in the reference configuration, respectively. In general, zero lower index indicates quantity in the reference configuration throughout this paper. Following principle of stationary potential energy, for all $\delta\boldsymbol{u}: B_0 \to \mathbb{R}^3$ vanishing on $\partial B_{0d}$, $\delta \Pi = 0$ holds [36]. Thus

$$D\Pi(\boldsymbol{\phi}) \cdot \delta\boldsymbol{u} \equiv \frac{d}{dh} \Pi(\boldsymbol{\phi} + h\delta\boldsymbol{u})\Big|_{h=0} = 0, \tag{2}$$

$$\delta\Pi = \int_{B_0} \boldsymbol{P} : \nabla_0 \delta\boldsymbol{u} \, dX - \int_{B_0} \delta\boldsymbol{u}(\boldsymbol{X}) \cdot \rho_0 \boldsymbol{b}_0(\boldsymbol{X}) dX - \int_{\partial B_0} \boldsymbol{T}_0(\boldsymbol{X}) \cdot \delta\boldsymbol{u}(\boldsymbol{X}) \, dS_0 = 0, \tag{3}$$

where $\nabla_0$ is the gradient with respect to material coordinates, and $\boldsymbol{P}$ is the first Piola-Kirchhoff (PK-I) stress tensor.

Next, suppose that a surface $\Gamma_0(\boldsymbol{X})$ with an arbitrary location within the body divides it into two regions $B_{01}$ and $B_{02}$, where $B_0 = B_{01} \cup B_{02}$. New surfaces (boundaries) are produced by the separation, such as $\Gamma_{01} \subset \partial B_{01}$ and $\Gamma_{02} \subset \partial B_{02}$. $\Gamma_{01}$ and $\Gamma_{02}$ are called inner boundary surfaces, on which inherently no external tractions are imposed. The separation is imaginary, not physically. By this, we mean that the matter is not separated like a fracture. In fact, there are internal tractions between the inner surfaces. The regions are shown in Fig. 2 and separated by $h$ value, which technically approaches zero so that the two regions are matched. Thus, we have

$$\Gamma_0(X) = \Gamma_{01}(X) = \Gamma_{02}(X), \tag{4}$$

where $\Gamma_0 = \partial B_{01} \cap \partial B_{02}$.

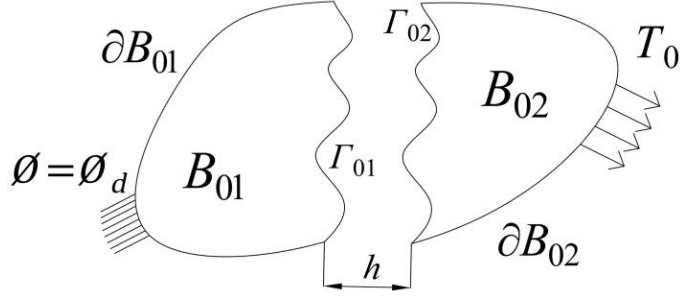

Fig. 2. The continuum body is split by the surface $\Gamma$.

Despite the equality between the inner surfaces, let us loosely define the relation between $\Gamma_1$ and $\Gamma_2$ as follows:

$$\nabla_0 \Gamma_1 - \nabla_0 \Gamma_2 = 0, \tag{5}$$

with $\Gamma_{02} = \Gamma_{01}$.

Integrating Eq. (5) gives

$$\Gamma_2 = \Gamma_1 + C, \tag{6}$$

therefore, for every $x_1 \in \Gamma_1$ and $x_2 \in \Gamma_2$, we have

$$x_2 = x_1 + c, \tag{7}$$

with $c_0 = 0$.

Equation (7) can be described as follows. The surface $\Gamma_2$ kinematically obeys the deformation of the surface $\Gamma_1$, and it also can translate with respect to the surface $\Gamma_1$. Given the relation between the inner surfaces, we start by decomposing the total potential energy as below

$$\widehat{\Pi} = \Pi_1 + \Pi_2 = \int_{B_{01}} W(F) \, dX - \int_{B_{01}} \boldsymbol{\phi}(X) \cdot \rho_0 \boldsymbol{b}_0(X) dX - \int_{\partial B_{01}} \boldsymbol{T}_0(X) \cdot \boldsymbol{\phi}(X) \, dS_0 + \int_{B_{02}} W(F) \, dX \tag{8}$$
$$- \int_{B_{02}} \boldsymbol{\phi}(X) \cdot \rho_0 \boldsymbol{b}_0(X) dX - \int_{\partial B_{02}} \boldsymbol{T}_0(X) \cdot \boldsymbol{\phi}(X) \, dS_0.$$

A "hat" over the total potential energy is used for emphasizing that this potential energy is not equal to the total potential energy described in Eq. (1). Considering the relation $\boldsymbol{P} = \partial W/\partial \boldsymbol{F}$ for hyperelastic materials, the variation of $\widehat{\Pi}$ can be written as

$$\delta \widehat{\Pi} = \int_{B_{01}} \boldsymbol{P} : \nabla_0 \delta \boldsymbol{u} \, dX - \int_{B_{01}} \delta \boldsymbol{u}(X) \cdot \rho_0 \boldsymbol{b}_0(X) dX - \int_{\partial B_{01}} \boldsymbol{T}_0(X) \cdot \delta \boldsymbol{u}(X) \, dS_0 + \int_{B_{02}} \boldsymbol{P} : \nabla_0 \delta \boldsymbol{u} \, dX$$
$$- \int_{B_{02}} \delta \boldsymbol{u}(X) \cdot \rho_0 \boldsymbol{b}_0(X) dX - \int_{\partial B_{02}} \boldsymbol{T}_0(X) \cdot \delta \boldsymbol{u}(X) \, dS_0. \tag{9}$$

Using the divergence theorem and separating the tractions on the surfaces $\Gamma_{01}$ and $\Gamma_{02}$, gives

$$\delta \hat{\Pi} = \int_{\partial B_{01} \backslash \Gamma_{01}} (\boldsymbol{P}\boldsymbol{n}_0).\delta \boldsymbol{u} \ dS_0 + \int_{\Gamma_{01}} (\boldsymbol{P}\boldsymbol{n}_0).\delta \boldsymbol{u} \ dS_0 - \int_{B_{01}} Div(\boldsymbol{P}).\delta \boldsymbol{u} \ dX - \int_{B_{01}} \delta \boldsymbol{u}.\rho_0 \boldsymbol{b}_0 dX$$

$$- \int_{\partial B_{01}} \boldsymbol{T}_0.\delta \boldsymbol{u} \ dS_0 + \int_{\partial B_{02} \backslash \Gamma_{02}} (\boldsymbol{P}\boldsymbol{n}_0).\delta \boldsymbol{u} \ dS_0 + \int_{\Gamma_{02}} (\boldsymbol{P}_2 \boldsymbol{n}_0).\delta \boldsymbol{u} \ dS_0 - \int_{B_{02}} Div(\boldsymbol{P}).\delta \boldsymbol{u} \ dX$$

$$- \int_{B_{01}} \delta \boldsymbol{u}.\rho_0 \boldsymbol{b}_0 dX - \int_{\partial B_{02}} \boldsymbol{T}_0.\delta \boldsymbol{u} \ dS_0, \tag{10}$$

where $\boldsymbol{n}_0$ is the outward unit normal of its surface. Substituting Eq. (7) into Eq. (10), considering Eq. (4), and noting that the $\boldsymbol{n}_0$ on $\Gamma_{01}$ is in the opposite direction of $\boldsymbol{n}_0$ on $\Gamma_{02}$, yields

$$\delta \hat{\Pi} = \int_{\partial B_{01} \backslash \Gamma_{01}} (\boldsymbol{P}\boldsymbol{n}_0).\delta \boldsymbol{u} \ dS_0 + \int_{\Gamma_{01}} (\boldsymbol{P}\boldsymbol{n}_0).\delta \boldsymbol{u} \ dS_0 - \int_{B_{01}} Div(\boldsymbol{P}).\delta \boldsymbol{u} \ dX - \int_{B_{01}} \delta \boldsymbol{u}.\rho_0 \boldsymbol{b}_0 dX \tag{11}$$

$$- \int_{\partial B_{01}} \boldsymbol{T}_0.\delta \boldsymbol{u} \ dS_0 + \int_{\partial B_{02} \backslash \Gamma_{02}} (\boldsymbol{P}\boldsymbol{n}_0).\delta \boldsymbol{u} \ dS_0 - \int_{\Gamma_{01}} (\boldsymbol{P}\boldsymbol{n}_0).\delta \boldsymbol{u} \ dS_0 + \int_{\Gamma_{02}} (\boldsymbol{P}\boldsymbol{n}_0).\delta \boldsymbol{c} \ dS_0$$

$$- \int_{B_{02}} Div(\boldsymbol{P}).\delta \boldsymbol{u} \ dX - \int_{B_{01}} \delta \boldsymbol{u}.\rho_0 \boldsymbol{b}_0 dX - \int_{\partial B_{02}} \boldsymbol{T}_0.\delta \boldsymbol{u} \ dS_0.$$

Applying the divergence theorem, noting that $(\partial B_{01} \backslash \Gamma_{01}) \cup (\partial B_{02} \backslash \Gamma_{02}) = \partial B_0$, and recalling the zero value of the external traction $\boldsymbol{T}_0$ at the inner boundary surfaces, Eq. (11) can be assembled to

$$\delta \hat{\Pi} = \int_{B_0} \boldsymbol{P}:\nabla_0 \delta \boldsymbol{u} \ dX - \int_{B_0} \delta \boldsymbol{u}(X).\rho_0 \boldsymbol{b}_0(X) dX - \int_{\partial B_0} \boldsymbol{T}_0(X).\delta \boldsymbol{u}(X) \ dX + \int_{\Gamma_{02}} (\boldsymbol{P}\boldsymbol{n}_0).\delta \boldsymbol{c} \ dS_0. \tag{12}$$

Equation (12) clearly breaks the continuity and Cauchy's lemma. This is because of our first assumption (Eq. (5)). For satisfying these principles, we must add $-\int_{\Gamma_{02}} (\boldsymbol{P}\boldsymbol{n}_0).\delta \boldsymbol{c} \ dS_0$ to Eq. (12), therefore the variation of the total potential energy modifies to

$$\delta \Pi = \delta \hat{\Pi} - \int_{\Gamma_{02}} (\boldsymbol{P}\boldsymbol{n}_0).\delta \boldsymbol{c} \ dS_0. \tag{13}$$

Equation (13) is exactly equal to the variation of the total potential energy for the non-partitioned body (Eq. (3)).

Remarks: Converting the interface $\Gamma$ into the inner surfaces $\Gamma_1$ and $\Gamma_2$ makes a problem for the relation between these two surfaces. In principle, they must be related to each other in order to reconstruct the non-partitioned continuum body. One option would be to link both surfaces rigidly so that they act as the unit interface $\Gamma$, i.e.,

$$\Gamma_2 = \Gamma_1 = \Gamma. \tag{14}$$

In other words, any deformation of the surface $\Gamma_2$ makes the same deformation of the surface $\Gamma_1$ and vice versa. This choice obviously returns us to the non-partitioned continuum body case. Alternatively, we assumed the deformation of the surface $\Gamma_{01}$ and the body $B_{02}$ are responsible for the deformation of the surface $\Gamma_{02}$ and instead of equalizing newborn surfaces, we equalized the gradients of the surfaces. However, as we have shown, this is not sufficient (Eq. (12)) since it could break the continuity principle, and hence the preventive force related to the translation of the surface $\Gamma_2$ has been added to restore the continuity. Adding this force is equivalent to fixing $\Gamma_2$ with respect to any deformation of the body $B_{02}$. As a result, another viable option is prepared by equalizing gradients of the surface $\Gamma_2$ with the surface $\Gamma_1$ along with fixing the surface $\Gamma_2$ with respect to any deformation of the body $B_{02}$, i.e.,

(a) $\nabla_0 \Gamma_2 = \nabla_0 \Gamma_1$,

(b) holding $\Gamma_2$ fixed with respect to any deformation of the body $B_{02}$. $\tag{15}$

Both constraint descriptions Eqs. (14) and (15) are equivalent to each other since they both return us to the non-partitioned continuum body case. The first description is a two-ways dependency ($\Gamma_2$ depends on $\Gamma_1$ and vice versa), but the second description, according to the constraint in Eq. (15), is shifted to a one-way dependency, i.e., only $\Gamma_1$ is able to drive $\Gamma_2$. As such, the surfaces $\Gamma_1$ and $\Gamma_2$ are called the principal and dependent surfaces. Obviously, it is possible to exchange the roles of the surfaces ($\Gamma_1$ as the dependent and $\Gamma_2$ as the principal surfaces) by exchanging them in Eq. (15) and changing $B_{02}$ to $B_{01}$. Concretely, the constraints in Eq. (15) may be described as follows. The deformation of $\Gamma_1$ makes the same deformation of $\Gamma_2$, therefore, $B_{01}$ affects $B_{02}$ through constraining displacement of $\Gamma_2$. On the other hand, internal traction on $\Gamma_2$ transmits to $\Gamma_1$ by the chain rule, and this is how $B_{02}$ affects $B_{01}$. The one-way dependency relation between the inner boundary surfaces, Eq. (15), forms the basis of all coupling methods that are presented in this paper.

## 3. Discretizing the partitioned continuum body

The finite element method (FEM) is widely accepted as a reliable and efficient computational tool for solving PDEs, especially in the solid mechanics. In this paper, this method is adopted for discretizing the continuum part, so the potential energy (Eq. (1)) in the concept of FEM can be written as

$$\widetilde{\Pi} = \sum_{el=1}^{n^{el}} \int_{B_0^{el}} W(\boldsymbol{F}^{el}(\sum_J N_J \mathbf{u}_J))dX - \sum_{I=1}^{n_{nodes}} \mathbf{f}_I^{ext} \cdot \mathbf{u}_I, \tag{16}$$

where $N_J$ is the shape function associated with the node $J$, $\boldsymbol{F}^{el}$ is the deformation gradient in the element $el$, $n^{el}$ is the number of elements, $n_{nodes}$ is the number of nodes, $\mathbf{u}$ and $\mathbf{f}$ are the nodal displacements and forces, respectively. The key parameter, which describes the material behavior, is the strain energy density $W$. Although the neo-Hookean model is extensively used for hyperelastic materials, we adopt a method known as local QC [12], in which the strain energy density is directly obtained via underlying atomistic energy. This sounds more natural for coupling atomistic region with continuum region as they can use the same potential; It is a good example of coarse-graining and is established on two fundamental assumptions. First, the Cauchy-Born rule, and second, atoms within an identical element experience the same atoms' environment during a deformation. Here, we consider simple crystals that can be described by the pair potentials. Following [12], the energy of every atom $i$ that belongs to an element can be written as

$$v^i(\{\boldsymbol{r}^{i,j}\}^i) = \sum_j v^{ij}, \qquad i \neq j, \tag{17}$$

in which $v^{ij}$ is the energy between the atoms $i$ and $j$, $\boldsymbol{r}^{i,j}$ is the relative position vector in the deformed configuration, and it is defined as the vector pointing from atom $i$ to atom $j$, i.e., $\boldsymbol{r}^{i,j} = \boldsymbol{r}^j - \boldsymbol{r}^i$, and the sequence $\{\boldsymbol{r}^{i,j}\}^i = \boldsymbol{r}^{i,1}, \boldsymbol{r}^{i,2}, \boldsymbol{r}^{i,3}, \ldots$ . To eliminate many degrees of freedom (dofs) for the purpose of coarse-graining, one needs to use the introduced assumptions. Turn to the first assumption, the Cauchy-Born rule for simple crystals states that the deformation of underlying atoms associated with a material point in a continuum body is equal to the deformation of the material point itself, i.e., it assumes the uniform deformation for the atoms beneath the material point. As a result, for two atoms $i$ and $j$ belong to the element $el$ by means of the Cauchy-Born rule, one may write $\boldsymbol{r}^{i,j} = \boldsymbol{F}^{el} \cdot \boldsymbol{R}^{i,j}$, where $\boldsymbol{R}^{i,j}$ is the relative position vector between atoms $i$ and $j$ in the reference configuration, i.e., $\boldsymbol{R}^{i,j} = \boldsymbol{R}^j - \boldsymbol{R}^i$. Therefore, the energy of each element is approximated by

$$v^{el} \approx \sum_{i=1}^{n_{atoms}^{el}} v^i(\boldsymbol{F}^{el}\{\boldsymbol{R}^{i,j}\}^i), \qquad i \neq j, \tag{18}$$

where $n_{atoms}^{el}$ is the number of atoms in the element $el$, and the sequence $\{\boldsymbol{R}^{i,j}\}^i = \boldsymbol{R}^{i,1}, \boldsymbol{R}^{i,2}, \boldsymbol{R}^{i,3}, \ldots$ . Satisfying the second assumption (atoms within an element experience the same atomic environment during the deformation) needs the deformation gradient to vary slightly by going through the neighbor elements of the element $el$. By imposing the second assumption, the energy of the element $el$ in Eq. (18) is approximated by

$$v^{el} \approx n^{el}_{atoms} v^{i^*}(\mathbf{F}^{el}\{\mathbf{R}^{i^*,j}\}^{i^*}), \qquad j = 1,2,3, \ldots \tag{19}$$

where $i^*$ is called the representative atom, and is located near the center of the element $el$. Once $v^{el}$ is determined, the strain energy density for each element may be calculated as

$$W^{el} = \frac{v^{el}}{n^{el}_{atoms}\Omega_0}, \tag{20}$$

where $\Omega_0$ is the volume of the Voronoi cell, and $n^{el}_{atoms}\Omega_0$ is the volume of the underlying atoms within the element.

Armed with the strain energy density, we now turn to the partitioned domain. We assume that the generated meshes for the body $B_{01}$ and the body $B_{02}$ are completely independent. As such, no elements interrupt the interface. In the partitioned body model, we have mentioned two inner surfaces $\Gamma_1$ and $\Gamma_2$. A question that might arise is which surface should be treated as a principal surface. The answer to the question according to our numerical results is given in Section 8. For now, let us say that $\Gamma_1$ is the principal surface. The total potential energy for the partitioned body can be written as

$$\widetilde{\Pi} = \sum_{el_1=1}^{n^{el_1}} \int_{B_{01}^{el_1}} W(\mathbf{F}(\sum_J N_J \mathbf{u}_J))dX + \sum_{el_2=1}^{n^{el_2}} \int_{B_{02}^{el_2}} W(\mathbf{F}(\sum_J N_J \mathbf{u}_J))dX - \sum_{I=1}^{n_{nodes}} \mathbf{f}_I^{ext} \cdot \mathbf{u}_I, \tag{21}$$

with the essential boundary condition $\mathbf{u} = \mathbf{u}_d$ on $B_{0u}$. Since $\Gamma_2$ is approximated by $\Gamma_1$, we have $d\Gamma_2 \approx d\Gamma_1$, and hence the constraint in Eq. (15-a) is satisfied. The constraint in Eq. (15-b) is also satisfied because $\Gamma_2$ is only depends on $\Gamma_1$. This is called the surface approximation method in the sense that we demand to approximate the inner surface $\Gamma_2$ for having the best match with the inner surface $\Gamma_1$ without any other further consideration. Undoubtedly, in ideal condition, $\Gamma_2$ must be equal to $\Gamma_1$ in a pointwise manner. Another approximation could be achieved by calculating force. Thus

$$-\frac{\partial \widetilde{\Pi}}{\partial \mathbf{u}^I} = -\sum_{el_1=1}^{n^{el_1}} \int_{B_{01}^{el_1}} \mathbf{P}:\frac{\partial \mathbf{F}}{\partial \mathbf{u}^I} dX - \sum_{el_2=1}^{n^{el_2}} \int_{B_{02}^{el_2}} \mathbf{P}:\frac{\partial \mathbf{F}}{\partial \mathbf{u}^I} dX + \int_{\Gamma_{01}} \mathbf{t}_1 dS_{01} + \int_{\Gamma_{02}} \mathbf{t}_2 dS_{02}, \tag{22}$$

with the essential boundary condition $\mathbf{u} = \mathbf{u}_d$ on $B_{0u}$, where $\mathbf{t}$ is the internal traction in the deformed configuration. The last two integrals must eliminate each other to satisfy the force equilibrium at the interface. Therefore, we invoke a condition in which $\Gamma_2$ is calculated in terms of $\Gamma_1$ in order to satisfy

$$\mathbf{t}_2 dS_{02} = -\mathbf{t}_1 dS_{01}. \tag{23}$$

Clearly, if Eq. (23) is satisfied, we have $\int_{\Gamma_{01}} \mathbf{t}_1 dS_{01} = -\int_{\Gamma_{02}} \mathbf{t}_2 dS_{02}$ in Eq. (22). Then, Eq. (23) can be rewritten based on PK-I:

$$\mathbf{P}_2 \mathbf{n}_{02} d\Gamma_{02} = -\mathbf{P}_1 \mathbf{n}_{01} d\Gamma_{01}. \tag{24}$$

Converting the PK-I stress tensor in Eq. (24) to Cauchy stress tensor $\boldsymbol{\sigma}$ gives

$$J_2 \boldsymbol{\sigma}_2 \mathbf{F}_2^{-T} \mathbf{n}_{02} d\Gamma_{02} = -J_1 \boldsymbol{\sigma}_1 \mathbf{F}_1^{-T} \mathbf{n}_{01} d\Gamma_{01}, \tag{25}$$

where $J$ is the Jacobian of the deformation gradient, and the superscript $T$ is the transpose operation. Using Nanson's formula ($J\mathbf{F}^{-T}\mathbf{n}_0 d\Gamma_0 = \mathbf{n} d\Gamma$), Eq. (25) leads to

$$\boldsymbol{\sigma}_2 \mathbf{n}_2 d\Gamma_2 = -\boldsymbol{\sigma}_1 \mathbf{n}_1 d\Gamma_1. \tag{26}$$

The condition in Eq. (15-b) is satisfied through the dependency of $\Gamma_2$ only on $\Gamma_1$, but the condition in Eq. (15-a) may or may not be satisfied. Because $d\Gamma_2 = d\Gamma_1$ is not the only solution to Eq. (26), the continuity may not be preserved. As a result, a geometrical consistency must be concerned when $\Gamma_2$ is approximated. The method is called force approximation in the sense that $\Gamma_2$ is approximated based on $\Gamma_1$ in order to produce matching forces at the interface (Eq. (23)). Considering the discretized body $B_{01}$ by elements and nodes, Eq. (23) can be written as

$$\mathbf{f}_2(X^\beta) = -\mathbf{f}_1(X^\beta), \quad \forall \beta \in \Gamma_1, \tag{27}$$

where $\mathbf{f}_1$ and $\mathbf{f}_2$ are the internal forces correspond to $\Gamma_{01}$ and $\Gamma_{02}$, respectively, and $\mathbf{f}(X^\beta)$ denotes the value of force at the node $\beta$. In summary, as we demonstrated, we need to approximate $\Gamma_2$ in terms of $\Gamma_1$ in order to not only satisfy Eq. (27) but also the geometrical consistency.

In this paper, we only concentrate on the constant strain FEM, for instance, tetrahedral element in 3D; as the body $B_{01}$ will be transformed into an atomistic model, the contribution of the continuum energy can be written as

$$\widetilde{\Pi} = \sum_{el_2=1}^{n^{el_2}} W(\mathbf{F}(\sum_J N_J \mathbf{u}_J)) V_0^{el_2} - \sum_{I=1}^{n_{nodes}} \mathbf{f}_I^{ext} \cdot \mathbf{u}_I, \tag{28}$$

where $V_0^{el_2}$ is the volume of the element $el_2$ in the reference configuration.

## 4. Coupling atomistic and continuum models: weak versus strong compatibility

Concurrent atomistic-continuum methods partition the body into two or more parts, each of which governs by its own theory. Indeed, the implementation of theories in their domains is straightforward, and the difficulty is how to compromise different theories through the interfaces. In Section 3, we have pointed out that the best compromising way is to impose $\Gamma_2 = \Gamma_1$ in a pointwise manner. This is called strong compatibility. Although the strong compatibility compromises the theories ideally, it has some drawbacks, such as requiring a specific mesh generation algorithm and having a relatively high computational cost. As a result, we have relaxed the strong compatibility ($\Gamma_2 \approx \Gamma_1$) as a surface or force approximation in Section 3. They are called weak compatibility. In this section, we present several coupling methods based on the surface and force approximation approaches for a multiscale model. First of all, let us transform $B_{01}$ from the continuum model to an atomistic model, which is considered to be more accurate, and we keep $B_{02}$ as the continuum model, which is discretized by finite elements (Fig. 3). Then, there are two approaches for coupling models. In the first approach, $\Gamma_1$ is selected as the principal surface, so the positions of the nodes on the surface $\Gamma_2$ are determined as a function of the positions of the atoms on the surface $\Gamma_1$ to achieve the force or surface approximation. As such, the interface atoms explicitly induce displacements to the interface nodes, and on the other side, forces on the interface nodes are transmitted to the interface atoms. Inversely, in the second approach, $\Gamma_2$ is selected as the principal surface.

Consider $\Gamma_1$ as the set of interface atoms and $\Gamma_2$ as the set of interface nodes, the strong compatibility coupling is achievable if and only if one-to-one correspondence relation exists between the atoms of $\Gamma_1$ with the nodes of $\Gamma_2$. Therefore, the number of interface nodes is equal to the number of interface atoms. The strong compatibility coupling is illustrated in Fig. 3 (a). The surface or force approximation becomes important when the number of nodes is less than the number of atoms at the interface (i.e., early mesh coarsening), as shown in Fig. 3 (b). Since $\Gamma_1$ belongs to the more accurate model (atomistic model), in all cases $\Gamma_1$ is considered to be the principal surface, i.e., $\Gamma_2 = \Gamma_2(\Gamma_1)$. The exception is the master-slave coupling method, which by its very definition $\Gamma_2$ is considered to be the principal surface, so we have $\Gamma_1 = \Gamma_1(\Gamma_2)$ in this case.

In this paper, we use Latin indices to indicate atoms and Greek indices to indicate nodes. Mathematically, the strong compatibility coupling is defined as a bijective function and can be written as

$$\forall i \in \Gamma_1, \exists! \ \alpha \in \Gamma_2 \ \text{and} \ \forall \alpha \in \Gamma_2, \exists! \ i \in \Gamma_1 \mid X^\alpha = R^i \ \rightarrow \ x^\alpha = r^i,$$
with
$$card(\Gamma_1) = N^{atoms}, \quad card(\Gamma_2) = N^{nodes}, \tag{29}$$

where the first statement (before the arrow) is the condition that the generated mesh needs to satisfy and the second statement (after the arrow) are the constraints must be imposed on nodes, $\exists!$ means ''there exists only,'' and $card$ stands for the cardinality of a set. In other words, for each atom, there exists one and only one node whose position is the same as its corresponding atom. Clearly, the total number of interface atoms is equal to the total number of interface nodes ($N^{atoms} = N^{nodes}$), and the condition in Eq. (29) constrains the mesh generation algorithm. The corresponding mesh to the strong compatibility, which satisfies the constraint in the Eq. (29), is called the ''fully-refined'' mesh.

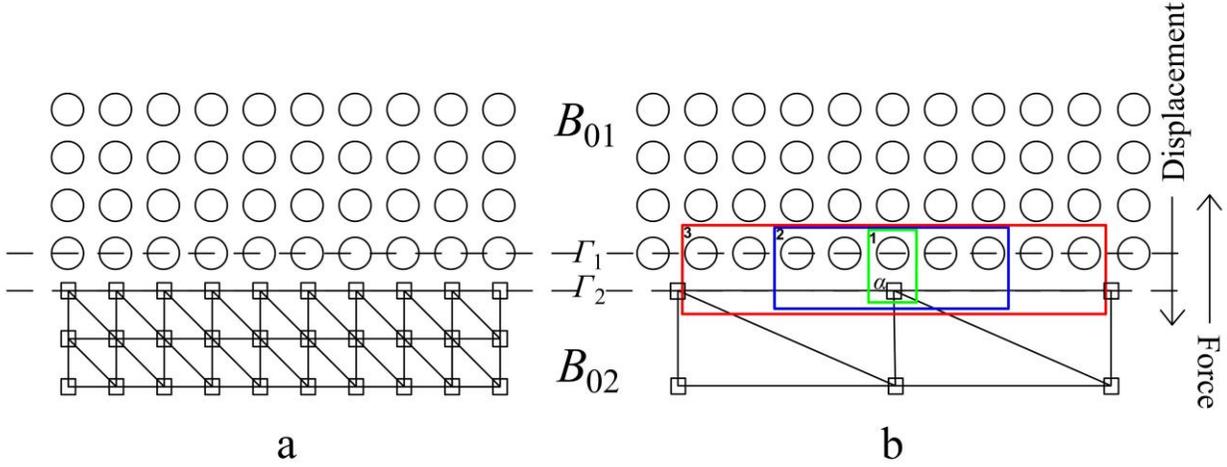

Fig. 3. Coupling strategies. The atoms and nodes are illustrated by circles and squares, respectively. The displacement and force arrows indicate that the interface atoms explicitly transmit displacement to the interface nodes, and in response, the interface nodes transmit force to the interface atoms. (a) Strong compatibility model with the fully-refined mesh. (b) Weak compatibility model with a coarse mesh. Associated atoms to the node $\alpha$ are indicated by the first, second, and third rectangles for the direct coupling, the $n^{th}$-nearest atoms, and the element-based schemes, respectively. Note that the surfaces $\Gamma_1$ and $\Gamma_2$ overlap, but here were drawn with a gap for more clarity.

## 4.1. Surface approximation approach

The surface approximation approach attempts to approximate the dependent inner surface to make the best fit with the principal surface. Either $\Gamma_1$ or $\Gamma_2$ can be selected as the dependent inner surface (and hence $\Gamma_2$ or $\Gamma_1$ will be the principal inner surface). Any interpolation or curve fitting methods can be used for approximating the dependent surface via the principal surface. Here, the direct and least-squares couplings are considered when $\Gamma_2$ is the dependent surface, and the master-slave coupling method is applied when $\Gamma_1$ is the dependent surface.

### 4.1.1 Direct coupling

One of the most useful interpolation methods is the polynomial-based one, whereby the dependent surface can be approximated. For example, the position of a typical interface node, say, $\alpha$ in Fig. 3(b), could be approximated by a group of nearby interface atoms that construct the local surface based on the appropriate polynomial. Now, consider a special case when every interface node coincides with the interface atoms. In this case, the polynomial interpolation constraint can be simplified to

$$\forall \alpha \in \Gamma_2, \exists!\ i \in \Gamma_1\ |\ X^\alpha = R^i\ \rightarrow\ x^\alpha = r^i,$$
with

$$card(\Gamma_1) = N^{atoms},\text{ and } card(\Gamma_2) = N^{nodes}. \tag{30}$$

The node $\alpha$ and its assigned atom (atom $i$) are shown in the first rectangle (green rectangle) in Fig. 3(b). In contrast to the strong compatibility, the condition in Eq. (30) allows $N^{nodes} < N^{atoms}$, which is necessary for early mesh coarsening at the interface. It is worth noting that under this condition, many interface atoms do not have any explicit contribution to constrain the interface nodes.

### 4.1.2 Least-squares coupling

The second method we consider to approximate $\Gamma_2$ is the least-squares method. The least-squares method was used as one of two compatibility conditions in the bridging scale method (BSM) and constrained the finite element nodes with atoms [7]. Here we apply the least-squares method within a concept of relaxing strong compatibility similar to Eq. (21). Thus, no iterative minimization is needed, unlike the BSM. The sum of squared residuals in the linear least-squares method can be expressed as

$$S_l = \sum_{i \in n_c}((\sum_{j=1}^{3} A_{lj}R_j^i) + B_l - u_l^i)^2, \quad l = 1,2,3. \tag{31}$$

where $n_c$ is the set of atoms is used for constraining node $\alpha$, $R^i$ is the position of atom $i$ in the reference configuration, $u^i$ is the displacement of atom $i$, and the lower indices in $R$ and $u$ indicate the component of the vector. Quantities **A** and **B** will be found as a function of atomic displacements by minimizing **S** with respect to **A** and **B**, i.e.,

$$\frac{\partial S}{\partial A} = 0, \quad \frac{\partial S}{\partial B} = 0. \tag{32}$$

Once **A** and **B** are determined, the positions of the interface nodes can be calculated as

$$x^\alpha = A(\{u^i\})X^\alpha + B(\{u^i\}) + X^\alpha, \tag{33}$$

in which $\{u^i\}$ is the set of displacements of atoms with $i \in n_c$. And we denotes the $card(n_c)$ by $n_{at}$. The set $n_c$ plays the main role in the method. Based on $n_c$, we define two schemes as follows. In the first scheme, $n_c$ is the set of first $n$ nearest atoms to node $\alpha$. In Fig. 3(b), the second rectangle (blue rectangle) shows $n_c$ when its cardinality ($n_{at}$) equals 5 (it can be equal to any other value). In the second scheme, $n_c$ is the set of projected atoms on elements that possess node $\alpha$. This scheme is shown in Fig. 3(b) with the third rectangle (red rectangle). Henceforth, the first and second schemes are abbreviated to the LS-$n_{at}$ and LS-EB schemes.

*4.1.3 Master-slave coupling*

The master-slave coupling method was originally used for coupling elements with different sizes in order to preserve continuity [37]. Later this approach was used for coupling finite elements with meshless methods [38]. Here we apply it for coupling the atomistic domain with the finite element domain. The master-slave coupling is able to completely match both inner surfaces, similar to the strong compatibility coupling, but with the key difference, the inner surface associated with the less accurate model ($\Gamma_2$) governs the inner surface associated with the more accurate model ($\Gamma_1$). As a result, the nodes of $\Gamma_2$ are considered master degrees of freedom and the atoms on the elements surfaces are considered slave degrees of freedom. The positions of the interface atoms update with the interface nodes through the shape functions. As such, the constraints can be written as

$$r^i = \sum_\alpha N^\alpha(R^i)x^\alpha, \quad \alpha \in \Gamma_2 \text{ and } i \in \Gamma_1 \tag{34}$$

where $N^\alpha(R^i)$ is the shape function defined at the node $\alpha$ and evaluated at the atom $i$. This method may represent the methods in which the relation between the inner surfaces is defined by $\Gamma_1 = \Gamma_1(\Gamma_2)$ for coupling domains.

4.2. Force approximation approach

The force approximation approach, in essence, approximates displacements of the interface nodes in order to make forces on the interface atoms get identical with the strong compatibility case. Here, we only concentrate on a specific form of function, in which the displacements of the interface nodes are approximated by averaging displacements of the interface atoms; hence we name it as consistent linear coupling (CLC) method. A general form of the average displacement for every $\alpha \in \Gamma_2$, can be written as

$$\mathbf{u}^\alpha = \frac{\sum_{i \in M_\alpha} c_i^\alpha \mathbf{u}_i}{\sum_{i \in M_\alpha} c_i^\alpha}, \tag{35}$$

where the set $M_\alpha \subsetneq \Gamma_1$ and will be defined subsequently, $\mathbf{u}_i$ is the displacement of atom $i$, $\mathbf{u}^\alpha$ is the displacement of node $\alpha$, $c_i^\alpha \geq 0$ and can be interpreted as an effect of atom $i$ on node $\alpha$. Accordingly, we write a logical constraint for the coefficients as below

$$\forall i,j \in M_\alpha \ for \ \|R^{i,\alpha}\| \leq \|R^{j,\alpha}\| \ we \ must \ have \ c_i^\alpha \leq c_j^\alpha, \tag{36}$$

where $R^{i,\alpha}$ is the relative position between atom $i$ and node $\alpha$ in the reference configuration ($X^\alpha - R^i$), similarly $R^{j,\alpha} = X^\alpha - R^j$, and $\|\blacksquare\|$ denotes the L2-norm. In other words, the effect of an atom with a long distance from node $\alpha$ cannot be more than an atom with a shorter distance. At this stage, the issue is how to determine $c_i^\alpha$ and $M_\alpha$. Once they are determined, $\mathbf{u}^\alpha$ will be known as well through Eq. (35). Based on the force approximation concept, $c_i^\alpha$ must be found so that the forces on the interface atoms be the same as the strong compatibility case. Therefore, the constraint in Eq. (27) can be rewritten as

$$\mathbf{f}_f(R^i) = \mathbf{f}_c(R^i), \quad \forall i \in \Gamma_1 \tag{37}$$

where lower indices $c$ and $f$ stand for a coarse and a fully-refined mesh, respectively, and $\mathbf{f}(R^i)$ denotes the force evaluated at the position of atom $i$. The minus sign in Eq. (27) is removed since both meshes are on one side (Fig. 4).

We present our formulation in an ideal case by noting that the size of the interface elements in the coarse mesh model is larger than the interface elements in the fully-refined mesh model. The ideal case is defined as an infinite flat interface with a "regular node distribution," and the interface elements are defined as the elements that at least have one node at the interface. By the term "regular node distribution," we mean that every node in the coarse mesh experiences the same atomic environment. An interface in the ideal case is illustrated in Fig. 4. All the interface nodes in the coarse mesh have the same atomic neighbors as node $\alpha$.

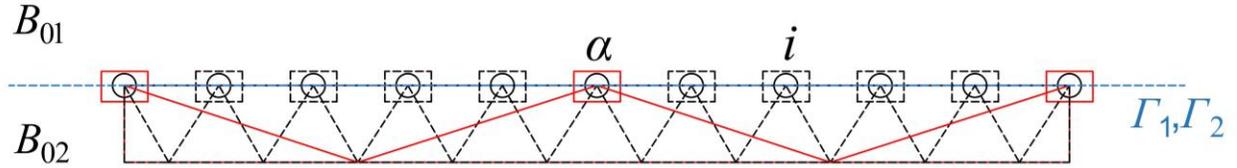

Fig. 4. Part of the fully-refined and coarse mesh models at the interface in the ideal case in 2D. Dashed black lines and solid red lines form the fully-refined mesh model and coarse mesh model, respectively. In the coarse mesh, $\alpha$ is a typical interface node, and $i$ is a typical interface atom. In the fully-refined mesh, $i$ is a typical interface node and interface atom at the same time, and $\alpha$ is not required due to playing the same role as $i$. Note that the regular atoms (atoms that are not interface atoms) in the body $B_{01}$ are not shown due to more clarity.

In general, the interface elements fill the space along the interface boundary for both mesh models (see Fig. 4). Thus,

$$\sum_{I=1}^{n_f^{el}} V_f^{el_I} = \sum_{J=1}^{n_c^{el}} V_c^{el_J}, \tag{38}$$

where $V_f^{el_I}$, $n_f^{el}$, $V_c^{el_J}$, and $n_c^{el}$ are the volume of the interface element $el_I$ in the fully-refined mesh, the number of the interface elements in the fully-refined mesh, the volume of the interface element $el_J$ in the coarse mesh, and the number of the interface elements in the coarse mesh, respectively. For the ideal case when the volumes are identical, Eq. (38) can be simplified to

$$n_f^{el} V_f = n_c^{el} V_c. \tag{39}$$

It might also be noted that the number of the interface elements is linearly proportional to the number of the interface nodes in both fully-refined and coarse mesh models:

$$n_f^{el} = k n^{atoms}, \tag{40a}$$

$$n_c^{el} = k n^{nodes}, \tag{40b}$$

where $k$ is a constant. For example, for the 2D model shown in Fig. (4), it is equal to 2 (first and last atoms or nodes are counted as half atoms or nodes, and the depth of the 2D model is a unit of length). Combining Eq. (39) with Eq. (40), gives

$$\frac{n^{atoms}}{n^{nodes}} = \frac{V_c}{V_f}. \tag{41}$$

In the ideal case, the regular node distribution at the interface implies that there are a certain number of atoms, which repeat periodically by each node in the coarse mesh model. Therefore, for each node $\alpha \in \Gamma_2$ we can assign certain atoms to node $\alpha$ based on their distances. Mathematically, it can be expressed as

$$S_\alpha = \{i| \quad \|\boldsymbol{R}^{i,\alpha}\| \leq \|\boldsymbol{R}^{i,\beta}\| \quad \forall \beta \in \Gamma_2, \forall i \in \Gamma_1, \alpha \neq \beta\}, \tag{42}$$

Then for each $i \in S_\alpha$ we define

$$Q_\alpha^i = \{\beta| \quad \|\boldsymbol{R}^{i,\alpha}\| - \|\boldsymbol{R}^{i,\beta}\| = 0, \forall \beta \in \Gamma_2, \alpha \neq \beta\} \cup \{\alpha\}. \tag{43}$$

From Eq. (43) we can capture an important quantity, namely the number of atoms per node, which is given by

$$n_\alpha^{atoms} = \sum_{i \in S_\alpha} \frac{1}{card(Q_\alpha^i)}. \tag{44}$$

Given the definition of $n_\alpha^{atoms}$, Eq. (41) can be rewritten as

$$\frac{1}{n_\alpha^{atoms}} = \frac{V_f}{V_c}. \tag{45}$$

For example, the number of atoms per node ($n_\alpha^{atoms}$) for Fig. 4 is equal to 5 by the use of Eq. (44), which is exactly equal to $V_f/V_c$. It is important to note that from the definitions of Eqs. (42), (43), and (44) sum of $n_\alpha^{atoms}$ on all interface nodes gives the total number of interface atoms, i.e.,

$$\sum_\alpha n_\alpha^{atoms} = N^{atoms}. \tag{46}$$

Then, consider the fully-refined mesh model shown in Fig. 4. The energy of elements from the discretized continuum side ($B_{02}$) contributes to the force on each atom/node $i \in \Gamma_1$ is equal to

$$E_{f,i}^{el} = \sum_J W_f^{el_J} V^{el_J}, \quad el_J \in El_f, \tag{47}$$

where $El_f$ is a set of elements, each of which possesses atom/node $i$. Considering the identical volume for elements (uniform mesh), the corresponding force on the atom/node $i$ can be calculated as

$$\boldsymbol{f}_f^i = -\frac{\partial E_{f,i}^{el}}{\partial \boldsymbol{u}_i} = -\sum_J \frac{\partial W_f^{el_J}}{\partial \boldsymbol{F}^{el_J}} \frac{\partial \boldsymbol{F}^{el_J}}{\partial \boldsymbol{u}_i} V^{el_J} = -V_f \sum_J \frac{\partial W_f^{el_J}}{\partial \boldsymbol{F}^{el_J}} \frac{\partial \boldsymbol{F}^{el_J}}{\partial \boldsymbol{u}_i}, \quad el_J \in El_f. \tag{48}$$

Similar to the fully-refined mesh model, in the coarse mesh model, the energy of elements from the discretized continuum side ($B_{02}$) contributes to the force on each atom/node $i \in \Gamma_1$ can be expressed as

$$E_{c,i}^{el} = \sum_J W_c^{el_J} V^{el_J}, \quad el_J \in El_c, \tag{49}$$

where $El_c$ is a set of elements containing nodes dependent on atom $i$ (recall that $\Gamma_2 = \Gamma_2(\Gamma_1)$). Note that $i$ for the coarse mesh model may not be a node, as can be seen in Fig. 4. Therefore, force on each atom $i$ at the interface may be calculated as

$$\mathbf{f}_c^i = -\frac{\partial E_{c,i}^{el}}{\partial \mathbf{u}_i} = -\sum_\alpha \frac{\partial E_{c,i}^{el}}{\partial \mathbf{u}_\alpha}\frac{\partial \mathbf{u}_\alpha}{\partial \mathbf{u}_i} = -\sum_J V^{el_J} \sum_\alpha \frac{\partial W_c^{el_J}}{\partial \mathbf{F}^{el_J}}\frac{\partial \mathbf{F}^{el_J}}{\partial \mathbf{u}_\alpha}\frac{\partial \mathbf{u}_\alpha}{\partial \mathbf{u}_i}, \quad el_J \in El_c. \tag{50}$$

Substituting the average displacement formulation from Eq. (35) to Eq. (50), yields

$$\mathbf{f}_c^i = -\sum_J V^{el_J} \sum_\alpha \frac{\partial W_c^{el_J}}{\partial \mathbf{F}^{el_J}}\frac{\partial \mathbf{F}^{el_J}}{\partial \mathbf{u}_\alpha} \frac{c_i^\alpha}{\sum_{i \in M_\alpha} c_i^\alpha}, \quad el_J \in El_c. \tag{51}$$

For the uniform mesh, and by changing the order of summations, Eq. (51) simplifies to

$$\mathbf{f}_c^i = -V_c \sum_\alpha \frac{c_i^\alpha}{\sum_{i \in M_\alpha} c_i^\alpha} \sum_J \frac{\partial W_c^{el_J}}{\partial \mathbf{F}^{el_J}}\frac{\partial \mathbf{F}^{el_J}}{\partial \mathbf{u}_\alpha}, \quad el_J \in El_c. \tag{52}$$

In Eq. (52), the summation for the coarse mesh is equivalent to the summation for the fully-refined mesh for sufficiently slow varying deformation gradient. Thus

$$\mathbf{f}_c^i \equiv -V_c \sum_\alpha \frac{c_i^\alpha}{\sum_{i \in M_\alpha} c_i^\alpha} \sum_J \frac{\partial W_f^{el_J}}{\partial \mathbf{F}^{el_J}}\frac{\partial \mathbf{F}^{el_J}}{\partial \mathbf{u}_i}, \quad el_J \in El_f. \tag{53}$$

Under the assumption of the regular node distribution, the value of $\sum_{i \in M_\alpha} c_i^\alpha$ is independent of $\alpha$, so we have

$$\sum_\alpha \frac{c_i^\alpha}{\sum_{i \in M_\alpha} c_i^\alpha} = \frac{\sum_\alpha c_i^\alpha}{\sum_{i \in M_\alpha} c_i^\alpha}. \tag{54}$$

To satisfy Eq. (37), by equalizing Eq. (48) with Eq. (53) and by using Eqs. (45) and (54), we have

$$\frac{\sum_\alpha c_i^\alpha}{\sum_{i \in M_\alpha} c_i^\alpha} = \frac{1}{n_\alpha^{atoms}}, \tag{55}$$

in which $n_\alpha^{atoms}$ must satisfy Eq. (46). Thus

$$\sum_\alpha \frac{\sum_{i \in M_\alpha} c_i^\alpha}{\sum_\alpha c_i^\alpha} = \frac{\sum_\alpha \sum_{i \in M_\alpha} c_i^\alpha}{\sum_\alpha c_i^\alpha} = N^{atoms}. \tag{56}$$

By changing the order of the summations, we have

$$\frac{\sum_{i \in \cup_\alpha M_\alpha} \sum_\alpha c_i^\alpha}{\sum_\alpha c_i^\alpha} = N^{atoms} \tag{57}$$

Eq. (57) can be rearranged as

$$\sum_{i \in \cup_\alpha M_\alpha} \sum_\alpha c_i^\alpha = N^{atoms} \sum_\alpha c_i^\alpha \tag{58}$$

The fact that the right-hand side of Eq. (58) depends on $i$, but the left-hand side does not, implies that $\sum_\alpha c_i^\alpha$ is a constant, i.e.,

$$\sum_\alpha c_i^\alpha = C, \tag{59a}$$

where $C$ is a constant. Therefore Eq. (58) simplifies to $\sum_{i \in \cup_\alpha M_\alpha} 1 = N^{atoms}$. Substituting Eq. (59a) into Eq. (55) yields

$$\sum_{i \in M_\alpha} c_i^\alpha = C n_\alpha^{atoms}. \qquad (59b)$$

By rescaling formulations, $c'^\alpha_i = c_i^\alpha / C$, Eqs. (59a) and (59b) becomes

$$\sum_\alpha c'^\alpha_i = 1, \qquad \textbf{I}$$

$$\sum_{i \in M_\alpha} c'^\alpha_i = n_\alpha^{atoms} \qquad \textbf{II}$$

which means $0 \leq c'^\alpha_i \leq 1$. Equations **I** and **II** are called consistency of force conditions. Another condition known as geometrical consistency needs to satisfy as we have demonstrated in the case of Eq. (22). It can be defined as

$$\sum_{i \in M_\alpha} c'^\alpha_i \boldsymbol{r}^{\alpha i} = \boldsymbol{0}. \qquad \textbf{III}$$

Strictly speaking, by satisfying conditions **I**, **II**, and **III** along with slow varying deformation gradient at the interface, we exactly reproduce the strong compatibility. However, we have yet to define the set $M_\alpha$ and coefficients $c'^\alpha_i$. Assuredly, they need to be defined in order to satisfy conditions **I**, **II**, **III**, and inequality in Eq. (36). Two schemes that meet these conditions are discussed as follows.

4.2.1 Atom-based scheme

By the term atom-based, we mean that classification is based on the interface atoms. First, the nearest node (say $\alpha$) to every atom (say $i$) is selected and assigned to the atom. Then, the coefficient $c'^\alpha_i$ is equalized to 1 unless the absolute value of relative distances between node $\alpha$ and other nodes with respect to the atom $i$ is less than a specified distance ($| \, \|\boldsymbol{R}^{i,\alpha}\| - \|\boldsymbol{R}^{i,\beta}\| \, | < h, \; \forall \beta \in \Gamma_2$). In that state, 1 is equally shared between all the corresponding nodes $\beta$'s and $\alpha$, and hence $c'^\alpha_i$ is determined. Mathematically, we define $M_\alpha$ and coefficients as below

$$M_\alpha = \{i \mid \alpha \in Q^i_\gamma, \forall i \in \Gamma_1, \forall \gamma \in \Gamma_2\} \qquad (60)$$

$$c'^\alpha_i = \frac{1}{card(Q^i_\alpha)} \times \begin{cases} 1, & i \in M_\alpha \\ 0, & i \notin M_\alpha \end{cases} \qquad (61)$$

If the specified distance approaches zero ($h = 0$), $M_\alpha$ becomes equivalent to the set $S_\alpha$, given by Eq. (42), and $c'^\alpha_i$ is related to the cardinality of the set $Q^\alpha_i$ as below

$$M_\alpha = S_\alpha \qquad (62)$$

$$c'^\alpha_i = \frac{1}{card(Q^i_\alpha)} \times \begin{cases} 1, & i \in M_\alpha \\ 0, & i \notin M_\alpha \end{cases} \qquad (63)$$

Since $M_\alpha$ is equal to $S_\alpha$, conditions **I** and **II** are satisfied automatically, i.e.

$$\sum_\alpha c'^\alpha_i = \sum_\alpha \frac{1}{card(Q^i_\alpha)} = 1 \qquad (64)$$

$$\sum_{i \in M_\alpha} c'^\alpha_i = \sum_{i \in M_\alpha} \frac{1}{card(Q^i_\alpha)} = \sum_{i \in S_\alpha} \frac{1}{card(Q^i_\alpha)} = n_\alpha^{atoms} \qquad (65)$$

Considering the choices of coefficients (Eq. (61)), the condition of geometrical consistency, which can be satisfied by interface nodes may be written as

$$\forall i \in M_\alpha \quad \boldsymbol{r}^{\alpha,i} = \boldsymbol{0} \; or \; \exists j \in M_\alpha \mid \boldsymbol{r}^{\alpha,i} + \boldsymbol{r}^{\alpha,j} = \boldsymbol{0}, \qquad (66)$$

which is a special case of the condition **III**.

In general, interface elements are large enough to keep the computational cost low but satisfy sufficient accuracy. Therefore, for each node, some atoms are always available to assign. This is the condition we defined $Q_\alpha^i$ in Eq. (43) for. We may change the definition of $Q_\alpha^i$ in order to increase the applicability of the method for conditions where distances between nodes at the interface are near or even lower than lattice constants. Thus, we replace the specified distance $h$ from zero to the distance between the nearest interface atoms to increase the flexibility of the method in Eq. (43), i.e.

$$Q_\alpha^i = \{\beta | \quad | \|\mathbf{R}^{i,\alpha}\| - \|\mathbf{R}^{i,\beta}\| | < \min\{\|\mathbf{R}^{m,n}\|\} \ \forall m, n \in \Gamma_1, \forall \beta \in \Gamma_2, \alpha \neq \beta\} \cup \{\alpha\}, \tag{67}$$

where $|\blacksquare|$ returns the absolute value. The coefficients and $M_\alpha$ are computed through Eqs. (60) and (61). The modification of $Q_\alpha^i$ allows us to reduce the size of interface elements near the lattice constant. For large elements at the interface, both definitions in Eqs. (43) and (67) are almost equal because most of the coefficients ($c_i^{\prime\alpha}$) are the same. However, by decreasing the size of the interface element, they could completely serve different solutions since, for the first definition, many nodes might be left without being constrained. Here, with the aim of unifying this scheme, we will only implement the definition given by Eq. (67) in Section 6. We refer to this scheme as an atom-based consistent linear coupling (CLC-AB) method hereinafter.

### 4.2.2 Element-based scheme

By the term element-based, we mean that determining the set $M_\alpha$ is based on the interface elements. In this scheme, the interface atoms whose projection on the inner surface $\Gamma_{02}$ are located beyond the interface elements that possess node $\alpha$ are considered with zero effect on the displacement of node $\alpha$. Atoms with non-zero effect constitute the set $M_\alpha$. Consequently, in contrast to the atom-based scheme, in general we have

$$M_\alpha \neq S_\alpha. \tag{68}$$

Let **P** be a projection operator that projects the interface atoms on triangles of the interface elements. We define the coefficients as

$$c_i^{\prime\alpha} = N^\alpha(\mathbf{P}(\mathbf{R}^i)). \tag{69}$$

We now claim that the coefficients defined by Eq. (69) satisfy the conditions **I**, **II**, and **III**.

The proof of satisfying condition **I** is straightforward. It is enough to consider the partitions of unity property of shape functions, which states $\sum_{\alpha=1}^{n^{nodes}} N^\alpha(\mathbf{X}) = 1$ for all $\mathbf{X}$, then we have

$$\sum_\alpha c_i^{\prime\alpha} = \sum_{\alpha=1}^{n^{nodes}} N^\alpha(\mathbf{P}(\mathbf{R}^i)) = 1. \tag{70}$$

We now turn to the second condition. Considering Eq. (69), the definition of $M_\alpha$, and the ideal case, for the purpose of converting the set $M$ to the set $S$ we can write

$$\sum_{i \in M_\alpha} c_i^{\prime\alpha} = \sum_{i \in M_\alpha} N^\alpha(\mathbf{P}(\mathbf{R}^i)) = \sum_{i \in S_\alpha} \frac{1}{card(Q_i^\alpha)} N^\alpha(\mathbf{P}(\mathbf{R}^i)) + \sum_{i \in \cup_\beta S_\beta} \frac{1}{card(Q_i^\beta)} N^\alpha(\mathbf{P}(\mathbf{R}^i)), \quad \alpha \neq \beta. \tag{71}$$

To explain the equality in Eq. (71), consider the assumption of the ideal case, which implicates that for an atom assigned to each node, there are atoms assigned to other nodes with the same relative positions. The latter atoms are the counterparts of the former atom. In Fig. 5, node $\alpha$ and its assigned atom (atom $j \in S_\alpha$) are shown. The counterparts of atom $j$ are $q$ and $p$, which are assigned to nodes $\beta_1$ and $\beta_2$, respectively, i.e., $q \in S_{\beta_1}$ and $p \in S_{\beta_2}$. Note that other counterparts of atom $j$ corresponding to nodes $\beta_3, \beta_4, \beta_5$, and $\beta_6$ are not located on the face of elements that possess node $\alpha$, hence not shown in Fig. 5. For enough large elements, we have $p, q \in M_\alpha$ whereas from the definition of $S_\alpha$ in Eq. (42), we have $q, p \notin S_\alpha$. Nonetheless, it is possible to consider $q$ and $p$ as the elements of the set $S$ associated with other neighbor nodes, e.g., $q \in S_{\beta_1}$ and $p \in S_{\beta_2}$. As a result, we may decompose the set $M_\alpha$ to the set $S_\alpha$ and $\cup_\beta S_\beta$, in which $\beta$ is the $\alpha$'s neighbor nodes. We added $1/card()$ in

Eq. (71) since, in contrast to the summation on the set $M$, atoms with equally spaced from the node $\alpha$ and the other nodes are calculated more than once in the case of summations on the set $S$. Next, the second summation on the right-hand side of Eq. (71) can be simplified to

$$\sum_{i \in \cup_\beta S_\beta} \frac{1}{card(Q_i^\beta)} N^\alpha(\mathbf{P}(\mathbf{R}^i)) = \sum_\beta \sum_{i \in S_\alpha} \frac{1}{card(Q_i^\alpha)} N^\beta(\mathbf{P}(\mathbf{R}^i)), \quad \alpha \neq \beta. \tag{72}$$

To explain the simplification, firstly, note that due to the ideal case we have $card(Q_i^\beta) = card(Q_i^\alpha)$. Secondly, the evaluated shape function at the projected position of an atom's counterparts ($q, p \in \cup_\beta S_\beta$) associated with the node $\alpha$ can be superseded by the evaluated shape function at the projected position of the atoms of the set $S_\alpha$ ($j \in S_\alpha$); however, associated with $\alpha$'s appropriate neighbor nodes. In our example, the neighbor nodes are $\beta_4$ and $\beta_5$ as can be seen in Fig. 5, so we have $N^\alpha(\mathbf{P}(r^q)) = N^{\beta_4}(\mathbf{P}(r^j))$ and $N^\alpha(\mathbf{P}(r^p)) = N^{\beta_5}(\mathbf{P}(r^j))$.

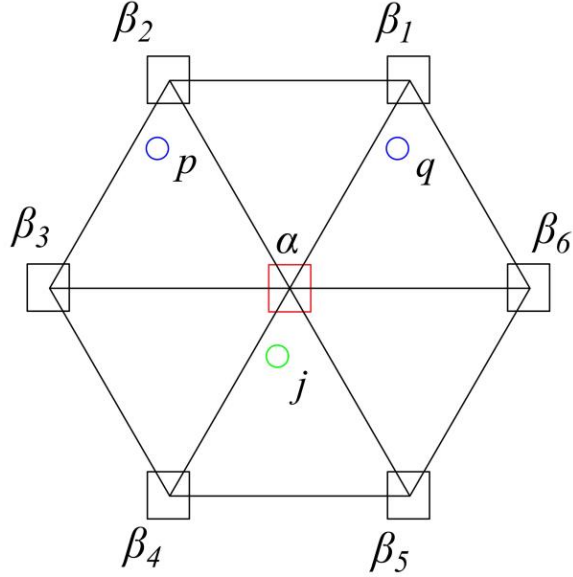

Fig. 5. Faces of the tetrahedral elements that possess a typical node $\alpha$ at the interface. Under the ideal case assumption, for a typical atom $j$ on the plane $\Gamma_1$, there must be atoms $p$ and $q$ on the plane $\Gamma_1$ as well.

Substituting Eq. (72) into Eq. (71) yields

$$\sum_{i \in S_\alpha} \frac{1}{card(Q_i^\alpha)} N^\alpha(\mathbf{P}(\mathbf{R}^i)) + \sum_\beta \sum_{i \in S_\alpha} \frac{1}{card(Q_i^\alpha)} N^\beta(\mathbf{P}(\mathbf{R}^i))$$
$$= \sum_{i \in S_\alpha} \frac{1}{card(Q_i^\alpha)} [N^\alpha(\mathbf{P}(\mathbf{R}^i)) + \sum_\beta N^\beta(\mathbf{P}(\mathbf{R}^i))], \quad \alpha \neq \beta \tag{73}$$

then, the partitions of unity property of shape functions for each node $\alpha$ requires that

$$N^\alpha(\mathbf{P}(\mathbf{R}^i)) + \sum_\beta N^\beta(\mathbf{P}(\mathbf{R}^i)) = 1, \quad \alpha \neq \beta. \tag{74}$$

Therefore, Eq. (73) leads to

$$\sum_{i \in S_\alpha} \frac{1}{card(Q_i^\alpha)} = n_\alpha^{atoms}. \tag{75}$$

Turning to the third condition, in terms of the element-based scheme, we will show that the ideal case assumption is enough for geometrical consistency. Here, we rename the nodes $\alpha$, $\beta_4$, and $\beta_5$ shown in Fig. 5, to 1, 2, and 3, respectively. As mentioned before, $p$ and $q$ are the counterparts of atom $j$. The condition **III** for node 1 is equal to

$$\sum_{i \in M_\alpha} c'^\alpha_i r^{i,\alpha} = c'^1_j r^{j,1} + c'^1_q r^{q,1} + c'^1_p r^{p,1}. \tag{76}$$

Then similar to simplification in Eq. (72), we produce an equivalent system, in which only a lower triangle from Fig. 5 is kept, and others are removed, as illustrated in Fig. 6. Therefore, in this system, rather than the counterparts of atom $j$ ($p$ and $q$), we deal with nodes 2 and 3, and hence Eq. (76) is equivalent to

$$c'^1_j r^{j,1} + c'^1_q r^{q,1} + c'^1_p r^{p,1} = c'^1_j r^{j,1} + c'^2_j r^{j,2} + c'^3_j r^{j,3} = N^1 r_1 + N^2 r_2 + N^3 r_3 \tag{77}$$

$$= \frac{A_{j23}}{A_{123}} r_1 + \frac{A_{j31}}{A_{123}} r_2 + \frac{A_{j12}}{A_{123}} r_3,$$

where $A$ is the area of a triangle and its lower index indicates its nodes/vertices, $r_1, r_2$, and $r_3$ are vectors initiate from atom $j$ to nodes, as shown in Fig. 6, and $N^1$, $N^2$, and $N^3$ are the shape functions evaluated at the position of atom $j$ and associated with nodes 1, 2, and 3, respectively. Further simplification of Eq. (77), takes it to zero as demonstrated below

$$\frac{1}{2A_{123}}(|det\ (r_1 \times r_2)|r_3 + |det\ (r_2 \times r_3)|r_1 + |det(r_1 \times r_3)|r_2)$$

$$= \frac{r_1 r_2 r_3}{2A_{123}} \big(\sin(\theta_{12})\ (\cos(\theta_3)\ i_1 + \sin(\theta_3)\ i_2) + \sin(\theta_{23})\ (\cos(\theta_1)\ i_1 + \sin(\theta_1)\ i_2)$$

$$+ \sin(\theta_{31})\ (\cos(\theta_2)\ i_1 + \sin(\theta_2)\ i_2)\big) = \mathbf{0}, \tag{78}$$

where $det$ stands for determinant, $\times$ is the cross product, $\theta_{ab} = \theta_b - \theta_a$, $i_1$ and $i_2$ are standard unit vectors, and $|\blacksquare|$ returns the absolute value. Thus, the three conditions are also satisfied for this scheme. We refer to this scheme as an element-based consistent linear coupling (CLC-EB) hereinafter.

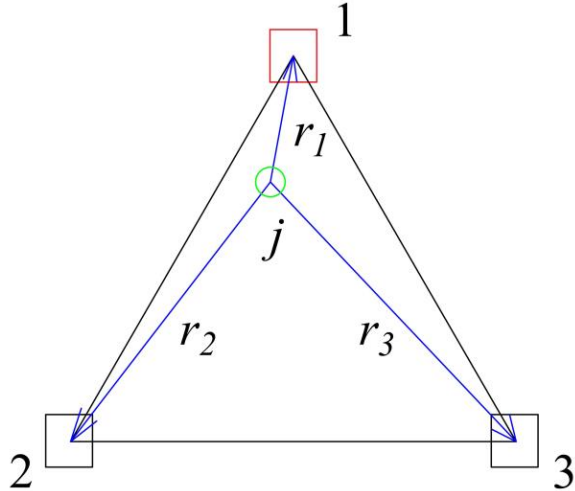

Fig. 6. The triangle with the vertices $\alpha, \beta_4,$ and $\beta_5$ from Fig. 5. The vertices are renamed as 1,2, and 3.

Even though the coefficients are determined in the ideal case for the CLC-AB and CLC-EB schemes, our numerical study will demonstrate that these schemes are accurate in spite of conditions where the ideal case is not exactly satisfied. Indeed, most of the mesh generation algorithms tend to produce a regular node distribution to reach a uniform mesh, and that provides sufficient accuracy for the CLC schemes. It is worth mentioning that even with a perfectly regular node distribution at the flat interface, the ideal case is inaccessible as the interface boundary is finite practically, i.e., at the edges and corners atomic environment for the interface nodes is not the same as the middle of the interface. This is a significant deviation from the ideal case, and therefore the highest errors are expected at the edges and corners. Theoretically, the CLC-AB scheme is less affected by deviation from the ideal case at the edges and corners than CLC-EB. This is because the coefficients of the CLC-AB are based on the set $S$, whereas the coefficients of the CLC-EB are based on the set $M$, which includes atoms with farther distances.

## 5. Energy expression for the atomistic-continuum body and finding the solution

The internal energy of the body could be calculated by the sum of the atomistic energy (domain $B_1$) with the continuum energy (domain $B_2$). However, a problem appears near the interface between the two regions since the energy is calculated in different ways, i.e., element-based in the domain $B_2$ and discrete-based in the domain $B_1$. Different approaches to calculating the energy are the cause of excessive counting of energy at the interface, known as double counting of energy. To have a closer look at the issue, consider a 2D lattice with a multiscale domain shown in Fig. 7, and assume that interactions between atoms can be described by the pair potential model, which is limited to the nearest-neighbor interaction (only interactions between the nearest neighbors are considered). The interface atoms are the atoms that are passed through by the inner surfaces $\varGamma_1$ and $\varGamma_2$, and the other atoms are called regular atoms. To illustrate the problem of the double counting of energy, consider the interface atoms $p$ and $q$, as shown in Fig. 7. The energy between these two atoms is treated atomically since they are atoms; however, half of this energy is computed by Eq. (28) through the element $el_1$ since they are nodes. As a result, the energy is calculated one and a half times more than the true energy. To overcome this difficulty, multiscale methods use different approaches, most of which decrease the contribution of the energy of the interface elements. For instance, the QC method employs Voronoi cells associated with the interface atoms and removes parts of the volumes of the interface elements that intersect with the volumes of the Voronoi cells [8]. The remedy we propose here is simple and reasonable. Rather than reducing the energies of the interface elements, reducing energies of the interface atoms are applied, i.e., the continuum energy is calculated by ignoring the atomistic part (like a fully continuum model), and modification is only applied to the atomistic energy. This approach ensures that the energy at the interface is exactly equal to the mono-scale model (fully atomistic or fully continuum model). By accepting the energy in the continuum region is entirely treated by the element-based approach without further modification, we conduct the discussion on how the energy of the atomistic domain is calculated. First of all, the energy between the regular atoms is entirely treated atomically. The condition which might be a little confusing is the energy between the interface atoms, which is discussed as follows. Consider the energy between the atoms $p$ and $q$ shown in Fig. 7. As half of this energy is calculated by the element $el_1$, the atomistic energy must be halved to give the true energy. Thus, the energy between atoms $p$ and $q$ is equal to

$$E^{pq} = \frac{1}{2}v^{pq} + E^{el_1}. \tag{79}$$

Now suppose that the regular atom $j$ shown in Fig. 7 transforms into the interface atom ($\varGamma_1$ and $\varGamma_2$ pass through atom $j$), therefore in this state element $el_1{'}$ with nodes $p$, $q$, and $j$ will be joined to the elements (a dashed triangle in Fig. 7). In this state, the energy between the interface atoms $p$ $and$ $q$ is calculated only by elements $el_1$ and $el_1{'}$.

$$E^{pq} = E^{el_1} + E^{el_1{'}}. \tag{80}$$

While we often encounter to the states that can be resolved by means of Eqs. (79) and (80), another state could be possible. For instance, consider two interface atoms $k$ and $m$ in Fig. 7. The energy between them is not taken into account by any elements. Therefore, the full atomistic energy must be considered in this state. Thus, we have

$$E^{km} = v^{km}. \tag{81}$$

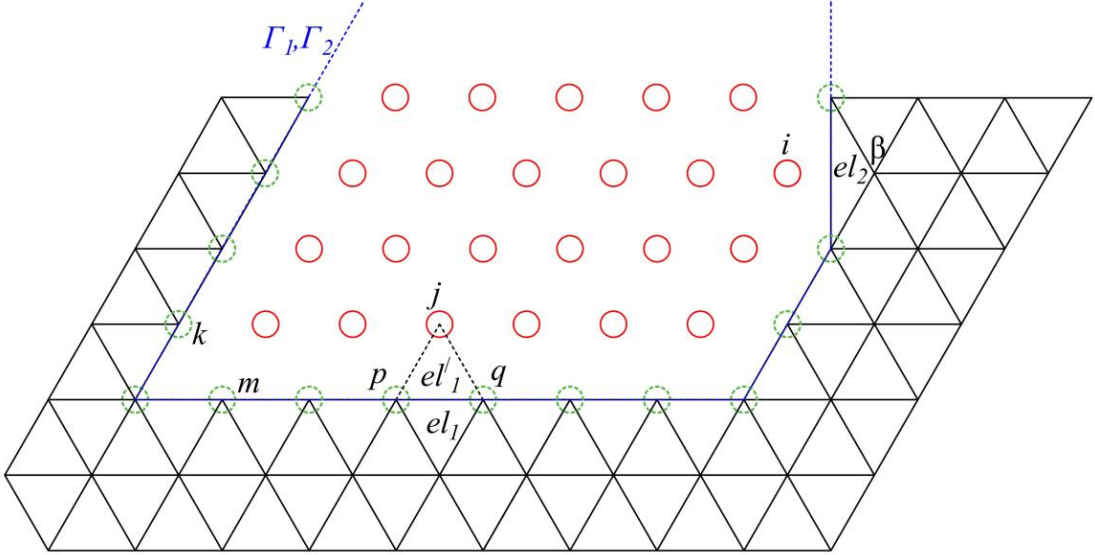

Fig. 7. Part of the two-dimensional multiscale body consists of elements and a triangular crystal lattice. Dashed green and solid red atoms are the interface and regular atoms, respectively. Note that the interface boundary is inappropriate at the right side near the atom $i$ since it makes inconsistency and must be modified by excluding the element $el_2$ or transforming the regular atom $i$ into the interface atom ($\Gamma_1$ and $\Gamma_2$ pass through atom $i$).

One important notification in terms of interface boundary must be added to complete our discussion regarding the consistency at interface. All of the regular atoms in Fig. 7 have full set of nearest neighbors (6 atoms) except atom $i$, this is because one nearest neighbor of atom $i$ is node $\beta$. This arise another problem since partial energy of atom $i$ and node $\beta$ is calculated by the element $el_2$ but no force associated with this energy exerts to atom $i$ due to locality property in continuum region. To overcome this difficulty, it is sufficient to modify the interface boundary by including atom $i$ or node $\beta$ as the interface atom. In other words, interface boundary cannot be arbitrary since all of the regular nodes must have their complete nearest neighbors either regular atoms or interface atoms. To ensure having a correct interface boundary, interface atoms should be constructed step-by-step with the nearest-neighbor atoms. Considering the above discussion, and recall that the energy in the coarse model is calculated without modifying, the total energy of the coupled atomistic-continuum domain can be written as

$$E = \hat{E}_{atomistic} + \tilde{\Pi} + E_{consistancy} + E_{external}, \tag{82}$$

where $\hat{E}_{atomistic}$ is the reduced atomistic energy as discussed above, $\tilde{\Pi}$ is the energy of the continuum part introduced by Eq. (28), $E_{consistancy}$ is the energy makes consistent force and energy for finite range interatomic interactions and the $E_{external}$ is the energy associated with external loads on atomistic domain. Determining $E_{consistancy}$ is postponed until Section 6.

Once the definition of the total energy is completed (Eq. (82)), it is need to find a solution by minimizing this energy functional. What one deals with at this point is a nonlinear nonconvex constrained optimization problem. In general, finding solution to these kind of problems are so difficult, but here the constraint equations are linear and homogeneous, which makes minimization simpler. All the variables in the multiscale problem are

$$\mathbf{u} = \{\mathbf{u}_I^{at}, \mathbf{u}_I^c, \mathbf{u}^o\}, \tag{83}$$

where $\mathbf{u}_I^{at}$ consists of the displacements of the interface atoms, $\mathbf{u}_I^{at}$ consists of the displacements of the interface nodes, and $\mathbf{u}^o$ consists of the displacements of the other nodes and atoms. Assembling all the interface displacements, we have

$$\mathbf{u}_I = \{\mathbf{u}_I^{at}, \mathbf{u}_I^c\}. \tag{84}$$

Then, the solution to the multiscale problem is

$$\mathbf{u}_{min} = \underset{\mathbf{u}}{argmin}(E), \tag{85}$$

s.t. $\mathbf{A}\mathbf{u}_I = \mathbf{0}$,

where $E$ is the total energy described in Eq. (82), $\mathbf{u}_{min}$ consists of the displacements, which minimize $E$, and $\mathbf{A}$ is a second-order tensor that makes the interface nodes and atoms relevant to each other and is given by the coupling methods, which we have discussed in Section 4. For example, the strong compatibility coupling (Eq. (29)), direct coupling (Eq. (30)), the least-squares coupling (Eq. (33)), the master-slave coupling (Eq. (34)), and the consistent linear coupling (Eq. (35)). The advantage of the introduced methods compared to the Lagrange-multiplier-based coupling methods is the simplicity of the constraint equations, which aids us to reformulate the problem only by certain variables. For all the methods discussed herein except master-slave coupling, variables are

$$\mathbf{u}^m = \{\mathbf{u}_I^{at}, \mathbf{u}^o\}, \tag{86}$$

but for the master-salve coupling, variables are

$$\mathbf{u}^m = \{\mathbf{u}_I^c, \mathbf{u}^o\}. \tag{87}$$

Therefore, Eq. (85) may reformulate as a nonlinear nonconvex optimization problem by considering $E$ only depends on $\mathbf{u}^m$, i.e.,

$$\mathbf{u}_{min}^m = \underset{\mathbf{u}^m}{argmin}(E). \tag{88}$$

## 6. Elastic nanoscale contact

In this section, the coupling methods introduced in Section 4 are compared, and a fully atomistic model is used as a benchmark. We describe a test model as follows. A diamond semi-sphere crystal and an aluminum face-centered crystal (FCC) are considered to be an indenter and a substrate, respectively. The geometry, dimensions, and crystal orientation of the intended problem are shown in Figs. 8 and 9, in which $a_0$ is the diamond lattice constant, $a_1$ is the aluminum lattice constant, $r_{c1}$ is the cutoff radius for interactions between the diamond and aluminum atoms, and all domains are symmetric with respect to the $X = 0$ and $Y = 0$ planes. Compared with the multiscale model, the fully atomistic model not only has the same geometry and dimensions but also has the same coordinates for all atoms that exist in the multiscale model. The fully atomistic model is illustrated in Fig. 10.

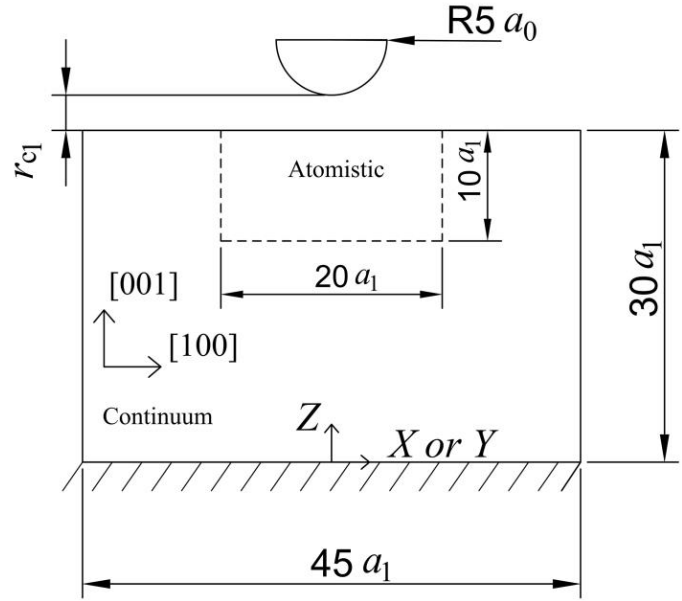

Fig. 8. The geometry and dimensions of the multiscale nanocontact model.

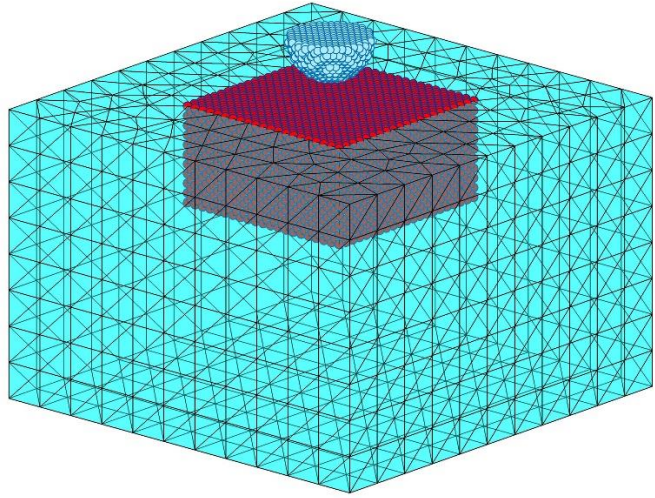

Fig. 9. The geometry of the multiscale nanocontact model in a 3D view.

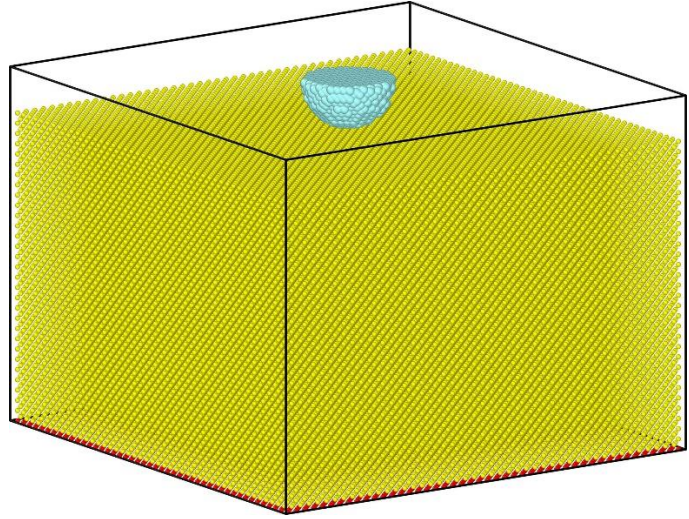

Fig. 10. The geometry of the fully atomistic nanocontact model [39]. The dimensions are exactly the same as in the multiscale model (for the substrate, the dimensions are $45\ a_1 \times 45\ a_1 \times 30\ a_1$, and the radius of the semi-sphere is $5\ a_0$).

To give a complete view of the test problem, we describe the boundary conditions, interatomic potential, and external potential as follows. The $X = -22.5\ a_1, X = 22.5\ a_1, Y = -22.5\ a_1,$ and $Y = 22.5\ a_1$ surfaces are set as the free surfaces, and atoms and nodes on the $Z = 0$ surface are held fixed during simulation for both fully atomistic (red atoms in Fig. 10) and multiscale models. The Lennard-Jones 12-6 potential is used for describing the interactions between the aluminum-aluminum atoms:

$$\phi(r) = 4\varepsilon \left[ \left(\frac{\sigma}{r}\right)^{12} - \left(\frac{\sigma}{r}\right)^{6} \right], \tag{89}$$

where $r$ is the distance between two atoms, $\varepsilon$ is the depth of the potential well, and $\sigma$ is the distance parameter. Usually, the cutoff radius $r_c$ is determined to reduce the computational burden and could be shortened to $1.5\ \sigma$ for L-J potential [40]. This is the so-called shifted forces cutoff at $1.5\ \sigma$, which ensures the continuity of the force and potential at the cutoff radius $r_{c2}$. Therefore, the potential energy is modified to

$$v(r) = \phi(r) - (r - r_{c2})\frac{d\phi}{dr}\Big|_{r=r_c} - \phi(r_{c2}), \quad r < r_{c2}$$

with $r_{c2} = 1.5\,\sigma$. (90)

The constants for aluminum suggested by [41] are $\varepsilon = 0.392175$ eV and $\sigma = 2.62$ Å. The assigned value to the cutoff radius ($r_{c2} = 1.5\,\sigma$), limits the interactions to the nearest neighbors for the aluminum crystal so that the simulations will be free of surface effects for both models and free of ghost forces for the multiscale model. Consequently, the accuracy and efficiency of the coupling methods will only be the center of attention. The absence of the ghost forces implies the consistency between the atomistic theory and the continuum theory near the interface, i.e., no additional effort is required for consistency, and therefore $E_{consistancy}$ is equal to zero in Eq. (82). The diamond semi-sphere is considered to be a rigid body, and the Morse potential is used for describing the interactions between the diamond and aluminum atoms [42]:

$$v^{ext}(r) = D_0\big[e^{-2\alpha(r-r_0)} - 2e^{-\alpha(r-r_0)}\big] \quad r < r_{c1}$$

$$D_0 = 0.28 \text{ eV}, \quad \alpha = 2.78 \text{ Å}^{-1}, \quad r_0 = 2.2 \text{ Å}, \quad r_{c1} = r_0 \tag{91}$$

As the nanoscale contact is assumed to be non-adhesive, the cutoff radius $r_{c1}$ in the Morse potential is equalized to $r_0$. Therefore, the net attractive force is neglected, i.e., the net force for this potential is always repulsive or neutral (equal to zero). Incidentally, the lattice constants for the diamond and aluminum crystals are considered to be $a_0 = 3.947$ Å and $a_1 = 4.254$ Å, respectively.

Returning to the total potential energy (Eq. (82)), $E_{consistancy}$ is equal to zero as discussed, $E_{external}$ is the energy between the aluminum and diamond atoms and is given by Eq. (91). The reduced atomistic energy ($\hat{E}$) is calculated such that energies between the interface atoms, marked by the red bonds in Fig. 11, are treated full atomistically, and only half of the energies between the other interface atoms are treated atomistically as the other halves are computed by the interface elements. It should be noted that all the regular atoms have the complete full-nearest neighbors, which is equal to 12 for the FCC crystal lattice. Therefore, the interface boundary has been implemented correctly.

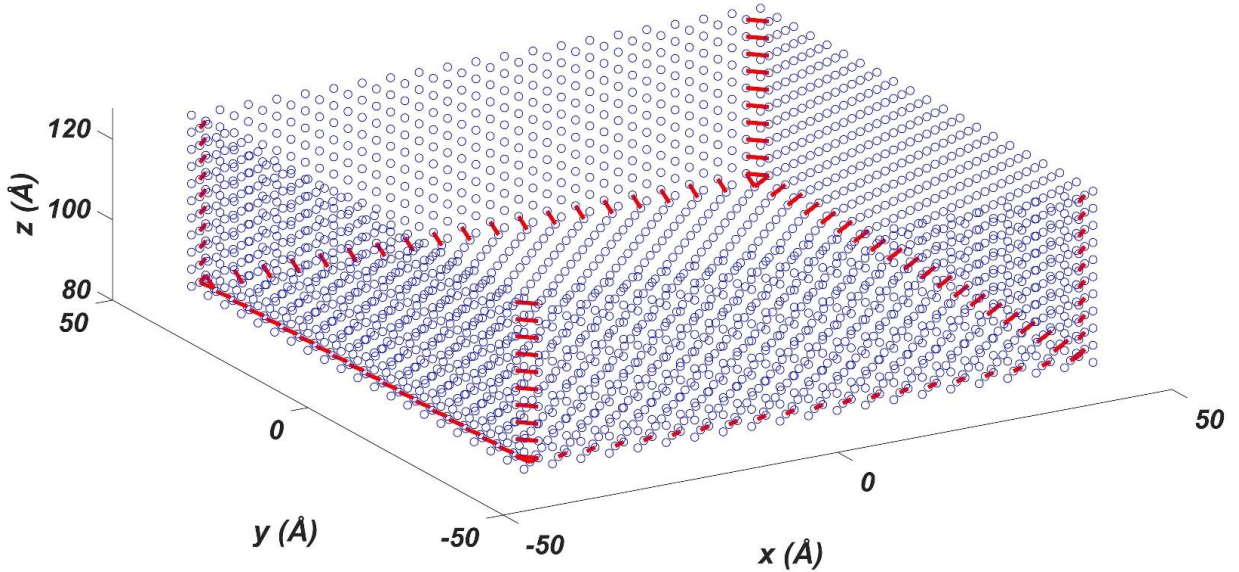

Fig. 11. Interface atoms. The energies of the atoms that share the red bond are treated full atomistically. The energies between the other interface atoms are treated half atomistically.

According to Fig. 8, no external interactions occur in the initial configuration since the minimum distance from the semi-sphere to the substrate is equal to the cutoff radius $r_{c1}$. Loading or unloading on the substrate happens when the diamond semi-sphere moves along the Z-axis. After its movement, the total potential energy is minimized for finding the solution (Eq. (88)). In this paper, in all simulations, the Broyden-Fletcher-Goldfarb-Shanno (BFGS) algorithm [43] is implemented to minimize the total potential energy introduced in Eq. (82). For the fully atomistic model, molecular statics (MS) simulation is performed by LAMMPS [44], and its solution is considered the exact solution. As we tested for the fully atomistic model, the contact remains elastic after 1 Å totally movement of the semi-sphere along the negative Z-direction in the sense that by inverse loading with the same increments, all atoms reach their initial positions. Before the presentation of the details and results of the simulations, we summarize the coupling methods introduced so far as

- **Strong compatibility coupling (SCC):** It represents the quasicontinuum (QC) or coupling of length scales (CLS) methods. Elements near the interface atoms must scale down so that the nodes coincide with atoms at the interface. The SCC needs a specific mesh, known as the "fully-refined (FR)" mesh.
- **Surface approximation approach (weak compatibility):** It approximates the discretized continuum inner surface to match with the atomistic inner surface, e.g., the direct coupling (DC), the least-squares schemes LS-$n_{at}$ and LS-EB. The exception is the master-slave coupling (MSC) method, in which the reverse approximation is applied.
- **Force approximation approach (weak compatibility)**: It approximates the discretized continuum inner surface to make forces on the interface atoms be the same as the strong compatibility model. The CLC-AB and CLC-EB have been developed in line with the force approximation approach.

To measure how much the methods are accurate, two main formulas are extensively used in the literature known as the global displacement percent error and energy percent error, defined by

$$Er_d^{CM} = \frac{\|\boldsymbol{u}^{MS} - \boldsymbol{u}^{CM}\|}{\|\boldsymbol{u}^{MS}\|} \times 100, \tag{92}$$

$$Er_E^{CM} = \left|\frac{\Delta E^{MS} - \Delta E^{CM}}{\Delta E^{MS}}\right| \times 100, \tag{93}$$

where $Er_d^{CM}$ is the global displacement error, $Er_E^{CM}$ is the energy error, and the upper indices $CM$ and $MS$ stand for the coupling method applied in the multiscale model, and molecular statics, respectively. Vectors in the global displacement error are assembled by displacements of certain atoms and are column vectors with $3M$ rows, where $M$ is equal to the number of substrate atoms in the multiscale model (see Table 1), and 3 is for the components of every displacement vector. $\boldsymbol{u}^{CM}$ is defined as a global displacement vector of the substrate atoms contained in the multiscale model. $\boldsymbol{u}^{MS}$ is the global displacement vector of the substrate atoms contained in the fully atomistic model (only atoms' counterparts in the multiscale model). The $L_2$-norm is the norm of the global displacement vector. For instance, $\|\boldsymbol{u}^{MS}\| = \sqrt{\sum_{m=1}^{3 \times M} u_m^2}$. In the energy error (Eq. (93)), "Δ" denotes the difference between the total energy of the current increment with the total energy of the previous increment. For the first increment, the previous total energy is considered when there is no interaction between the semi-sphere and the substrate. Finally, we emphasize that in all simulations, the geometry, boundary conditions, and minimization algorithm are exactly the same, and the only difference is in the coupling conditions.

To compare methods with weak compatibility, we produce a challenging condition by creating immense elements (type A mesh) where atoms meet nodes, as shown in Fig. 12. In this condition, the number of the interface atoms is equal to 2441, while only 96 nodes are available at the interface to couple with.

Table 1
Details of the fully atomistic and multiscale models for the aluminum substrate.

| Model | Mesh type | Number of atoms | Number of nodes | Number of fixed atoms | Number of fixed nodes | Number of elements | Atomistic region volume/total volume (%) | Dofs |
|---|---|---|---|---|---|---|---|---|
| Fully Atomistic | - | 252571 | - | 4141 | - | - | 100 | 745290 |
| Multiscale | FR | 17651 | 9520 | - | 144 | 47388 | 6.58 | 81081 |
| Multiscale | A | 17651 | 1415 | - | 144 | 6374 | 6.58 | 56766 |

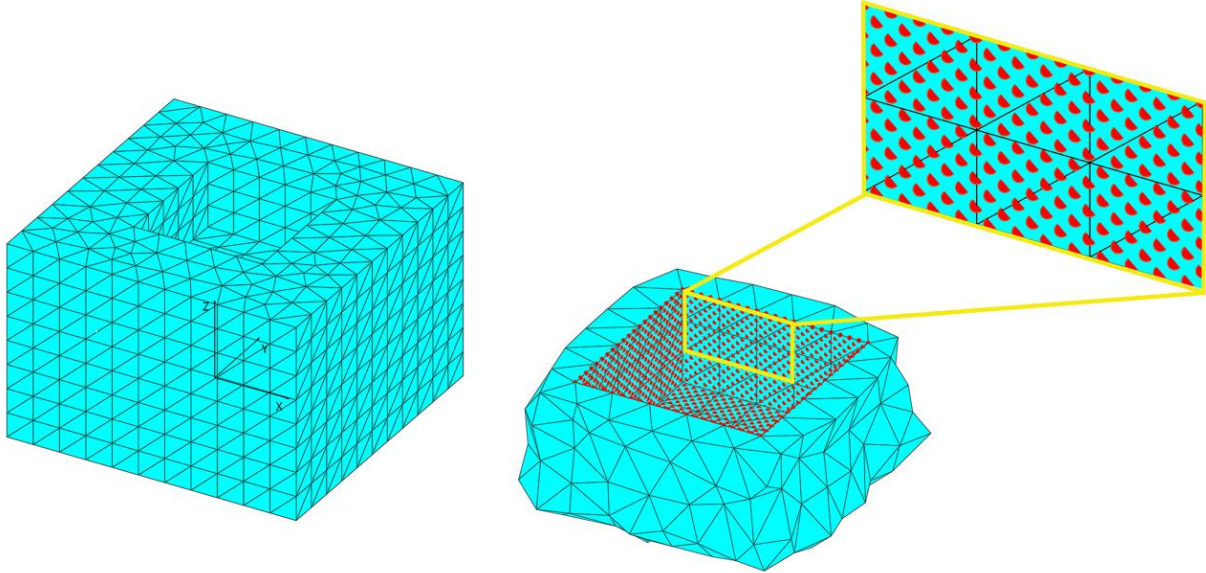

Fig. 12. (Left) Type A mesh. (Right) Interface atoms (red spheres) and interface elements of the type A mesh.

We first examine the coupling methods, which are based on the surface approximation approach, excluding the MSC method, by means of the global displacement error introduced in Eq. (92). Loading on the substrate is accomplished by only one increment defined by moving semi-sphere 0.1 Å along the negative Z-axis. We recall that $n_{at}$ is a critical parameter that determines the number of atoms that affect each node in the LS-$n_{at}$ scheme. Here the value of $n_{at}$ is considered fixed for every node, and its effect on the LS-$n_{at}$ scheme is investigated for all 5, 10, 20, 30, 40, 50, 60, 70, 80, 100, and 120 values. The global displacement errors for the DC, LS-EB, and LS-$n_{at}$ schemes are illustrated in Fig. 13.

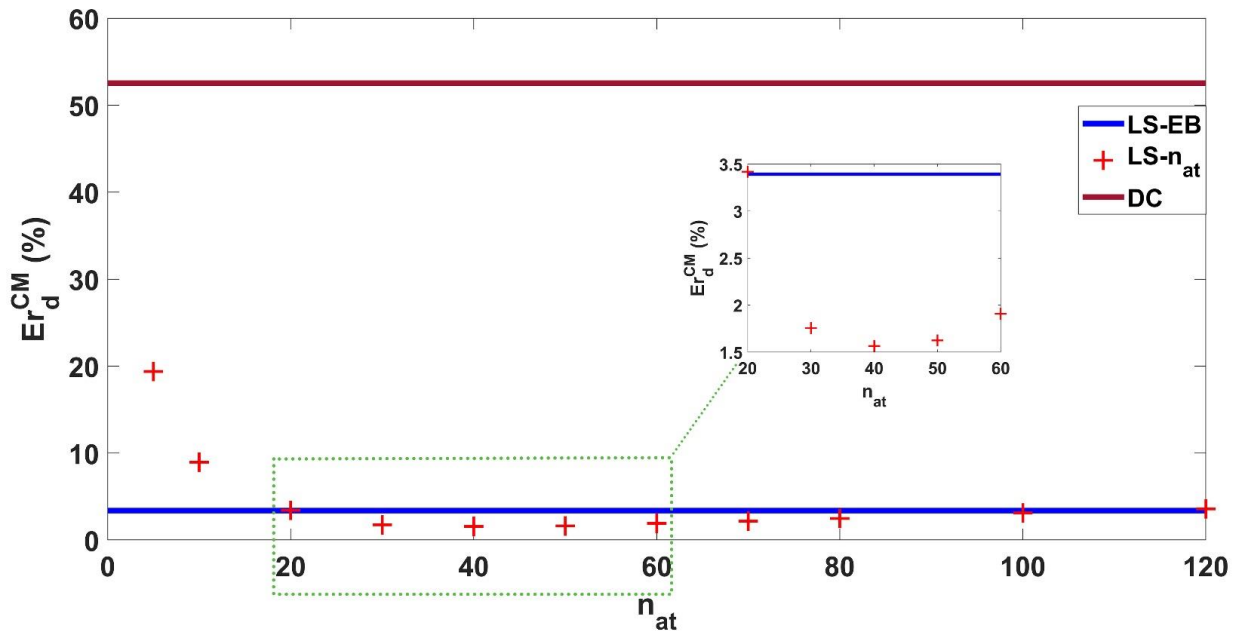

Fig. 13. Comparison of the global displacement error between the least-squares schemes and direct coupling.

The DC method shows upward of 50 percent error, making the method completely unusable. The LS-$n_{at}$ scheme does not have a unique solution as it depends on $n_{at}$. By increasing $n_{at}$ up to 40 atoms per node, the accuracy is increased as well; however, thereupon, the accuracy tends to decrease slightly. Moreover, choosing $n_{at}$ lower than 20, results in a huge error in the displacements and makes it comparable to DC. Similar to the DC method, the LS-EB scheme gives a unique solution by definition. However, the accuracy of the LS-EB is considerably better than the DC method. As can be seen in Fig. 13, all the LS-EB, LS-20, LS-100, and LS-120 have nearly the same accuracy. Accordingly, the LS-20 scheme is designated as a representative of the least-squares coupling. It is true that the LS-$n_{at}$ scheme has a potential to give a more accurate result, for instance, LS-40. However, the chosen $n_{at}$ for the best solution varied if the size of the elements near the interface would change. Therefore, further investigation is required to predetermine $n_{at}$.

We now compare all of the introduced coupling methods with each other. Once again, the type A mesh is employed to produce the challenging condition for the weak compatibility coupling methods. However, it is imperative that the SCC adopt the FR mesh. The type FR mesh is characterized by the coincidence of nodes with atoms at the interface, as can be seen in Fig. 14. The details of the type A mesh and the type FR mesh are listed in Table 1. For the sake of fair comparison regarding the computational cost between the strong compatibility and the weak compatibility couplings, the distribution of nodes on the boundary surfaces (clearly, except Z=30 $a_1$) are the same for their corresponding meshes. For example, the number of nodes on the surface Z=0 is equal to 144 for the type A and FR meshes, according to Table 1. Loading condition for all methods is the same and is imposed by 0.1 Å movement of the semi-sphere along the negative Z-axis for each increment to induce force on the substrate. Here we consider five increments; therefore, the semi-sphere totally moves 0.5 Å. After each incremental loading, the system is minimized for finding the solution. The global displacement errors (Eq. (92)) are plotted for the different coupling methods in Fig. 15.

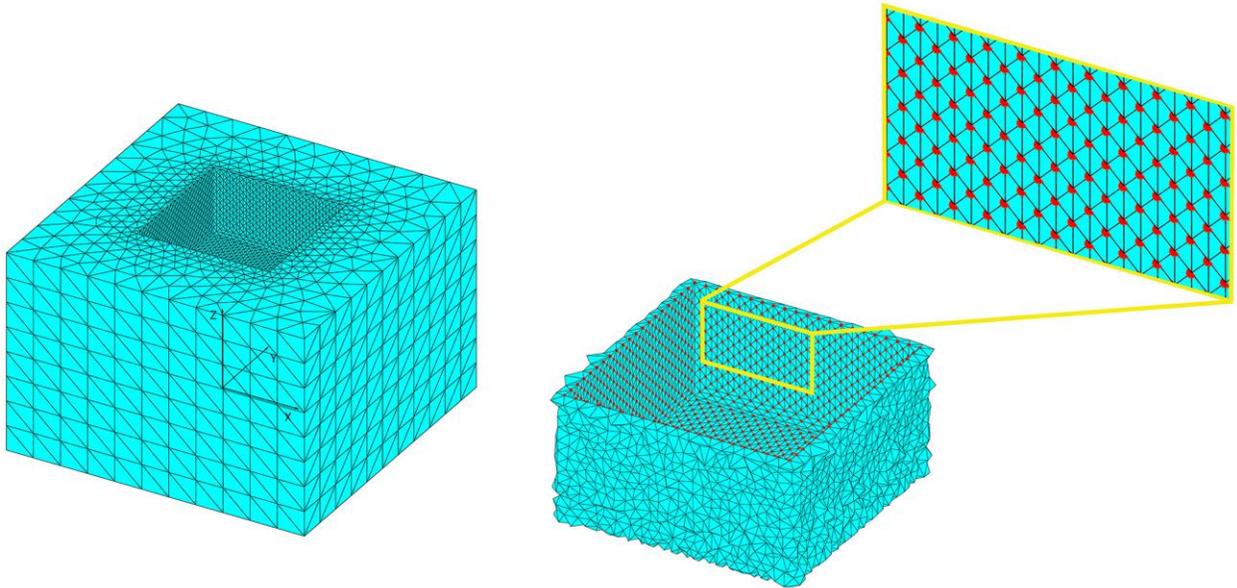

Fig. 14. (Left) Fully-refined mesh, and (right) its interface elements and atoms. The magnified view indicates that the interface atoms (red spheres) are coincident with the interface nodes.

According to Fig. 15, The SCC method, which is inherently equipped with the type FR mesh, has the least global displacement error. This was expected since SCC has significantly more dofs compared with the other coupling methods (see Table 1). However, the more computational cost is the consequence of this accuracy as will be apparent later on. Among the weak coupling methods equipped with the type A mesh (almost equal computational cost), the CLC-AB coupling method has the least error. The errors for the CLC-EB and CLC-AB schemes are close and are less than 2% (roughly twice the SCC). In the case of methods based on the surface approximation approach, the DC method has about 50 percent error (not shown in Fig. 15). The MSC method shows better accuracy compared to LS-20 (which represents the least-squares family), but we recall that LS-20 is not the best choice for the LS-$n_{at}$ scheme. Knowing that LS-40 is a better choice (see Fig. 13), the error for this scheme can be decreased down to the CLC-AB scheme as we tested but not shown in Fig. 15. The other measure of accuracy for coupling methods is the energy error introduced by Eq. (93). The energy error for various coupling methods are plotted in Fig. 16.

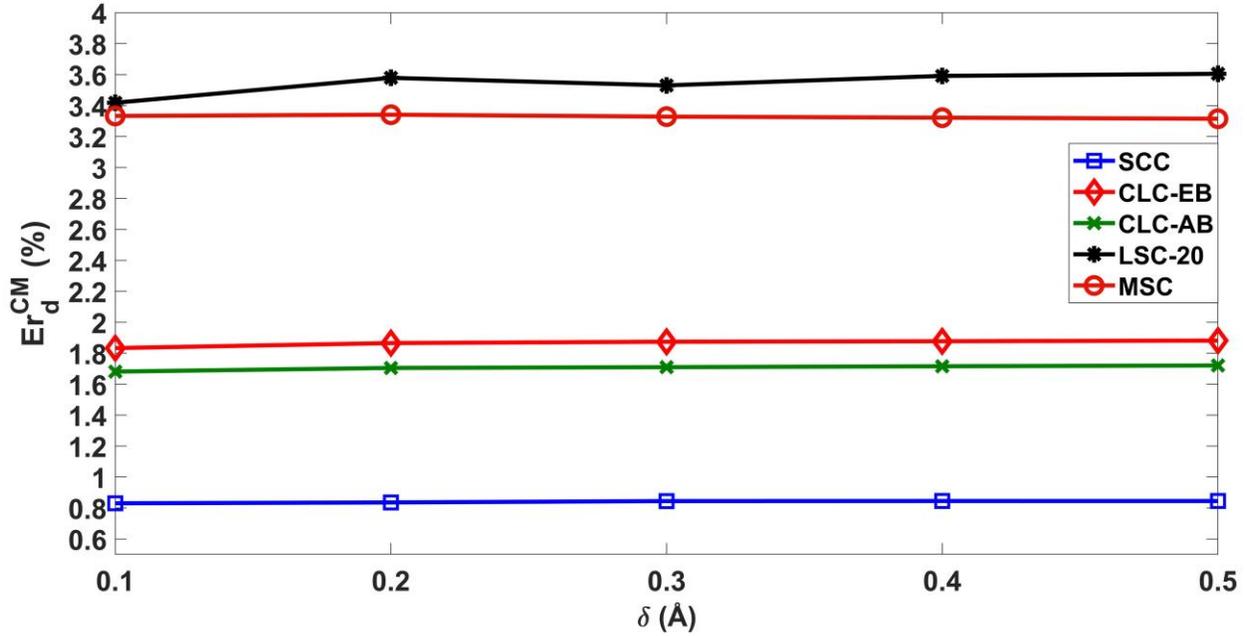

Fig. 15. Comparison of the global displacement error between the SCC and the other weak compatibility couplings.

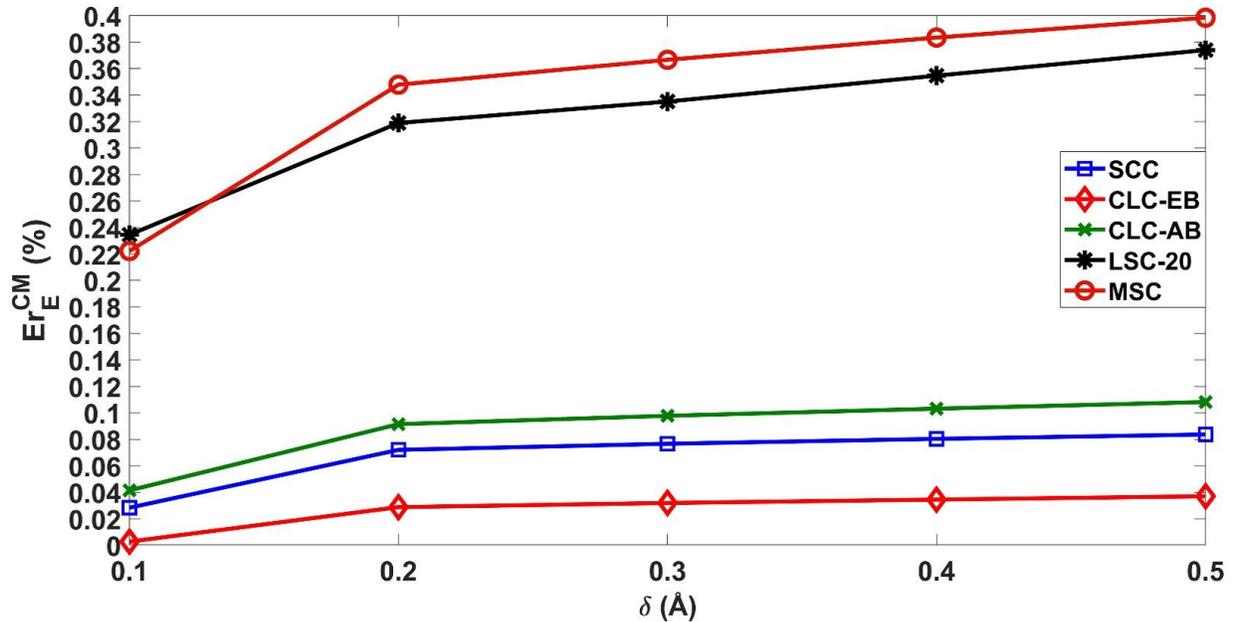

Fig. 16. Comparison of the energy error between the SCC and the other weak compatibility couplings.

According to Fig. 16, the CLC-EB scheme has the least energy error (even less than SCC). These results may not be expected from the results of the global displacement error plotted in Fig. 15. Because the total energy is computed based on the positions of atoms, it is supposed a method that finds the positions of atoms more precisely is the one with the less energy error. The reason why this statement seems false here stems from the fact that the global displacement error is the error on all substrate atomic displacements. On the other hand, the total energy consists of two parts: interatomic energy and external energy. Therefore, the statement could be true if the distribution of the absolute displacement error (the difference between the displacement of the atom in the multiscale model and the fully atomistic model) was uniform. Thus, after imposing the fifth increment, we calculate the norm of absolute displacement error for every atom in the substrate to shed light on the subject. Results are shown in Figs. 17-19 for the introduced methods. The error is uniform for none of which. This enables us to deduce in the CLC-EB scheme positions of a group of aluminum atoms that interact with the diamond atoms (contributors to the external potential energy) are found more accurately than in the other methods, and since the variation in the external potential is dominant (rather than interatomic potential), the error in the total energy is the least for CLC-EB.

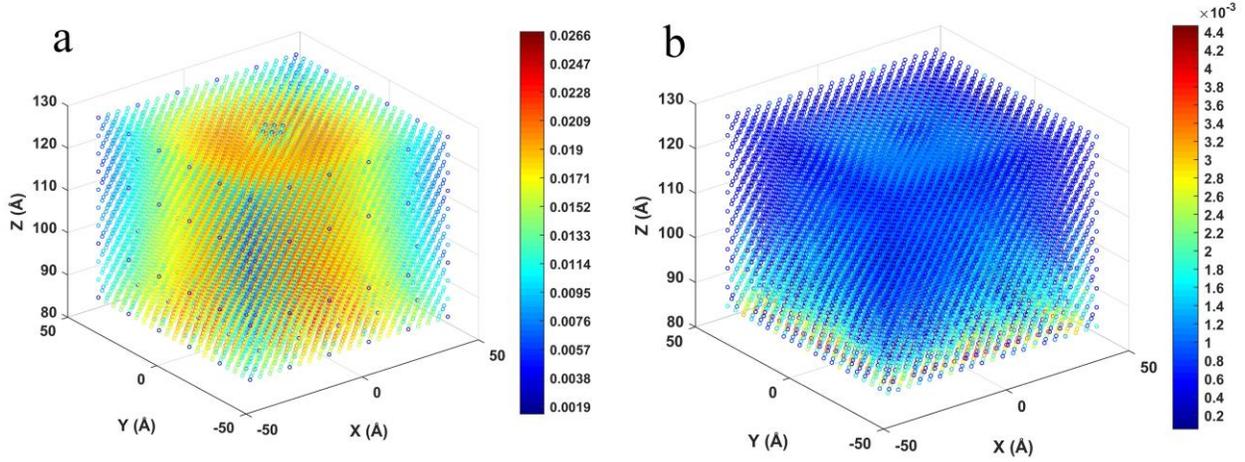

Fig. 17. Norm of absolute displacements error of the substrate atoms for the (a) DC and (b) LS-20 schemes.

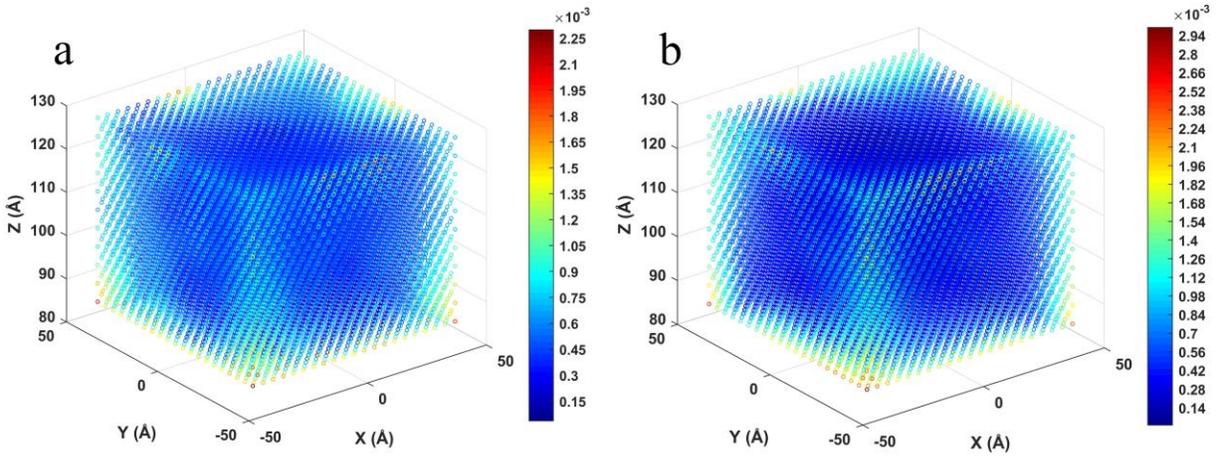

Fig. 18. Norm of absolute displacements error of the substrate atoms for the (a) CLC-AB and (b) CLC-EB schemes.

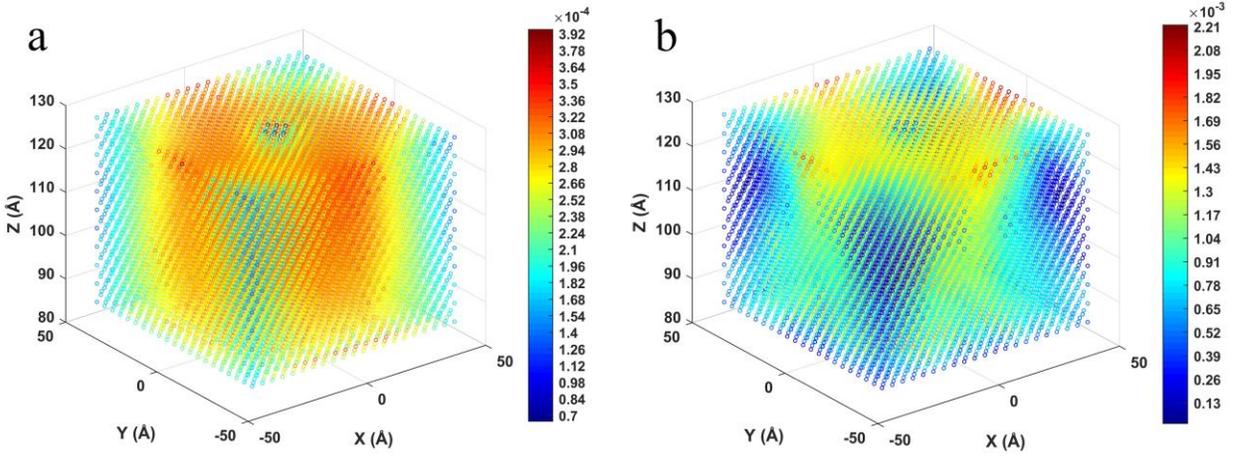

Fig. 19. Norm of absolute displacements error of the substrate atoms for the (a) SCC and (b) MSC methods.

To support our deduction, we compare the methods for two other subjects: (1) the absolute displacement error of the most displaced atom, plotted in Fig. 20, and (2) the semi-sphere reaction force error in the Z-direction, plotted in Fig. 21. The least errors for the CLC-EB scheme in Figs. 20 and 21, indicate that the positions of the substrate atoms underneath the diamond atoms are found more accurately for this method compared with the other methods. We now return to the distribution of the absolute displacement errors (Figs. 17-19), which might be more insightful. As pointed out by the theory in Section 4, the CLC method is practically less accurate at edges and corners due to deviation from the ideal case and lack of geometrical consistency. This is clearly can be seen in Fig. 18, in which corners and edges indicate the highest errors; however, by traveling toward the center of surfaces, errors are getting lower and lower. Furthermore, the errors of the corners and edges for CLC-AB are less than CLC-EB. This is why CLC-AB shows a lower global displacement error in Fig. 15. Another notable point is related to the similarity of the distribution of the absolute displacement error between methods. The CLC-AB and CLC-EB schemes have a similar distribution of error. Likewise, the MSC is analogous to SCC since they both preserve continuity completely. The LS method serves an entirely different distribution of error as the errors are nearly uniform on the upper layers and vary dramatically by moving toward the first few layers.

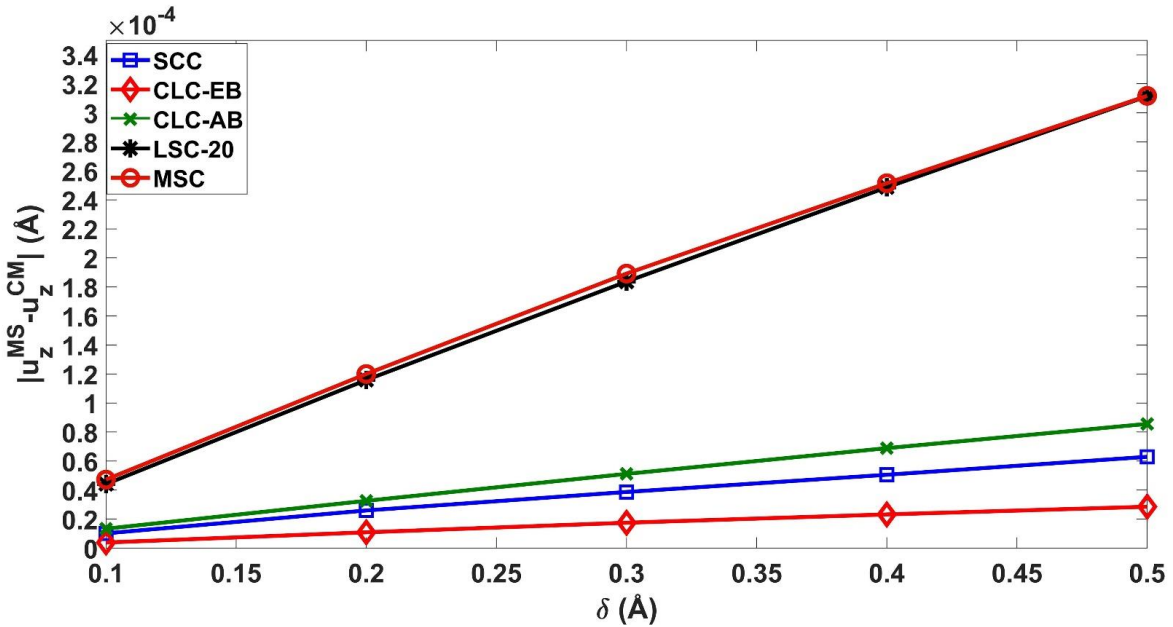

Fig 20. Comparison of the absolute displacement error for the most displaced atom between the SCC and the other weak compatibility couplings.

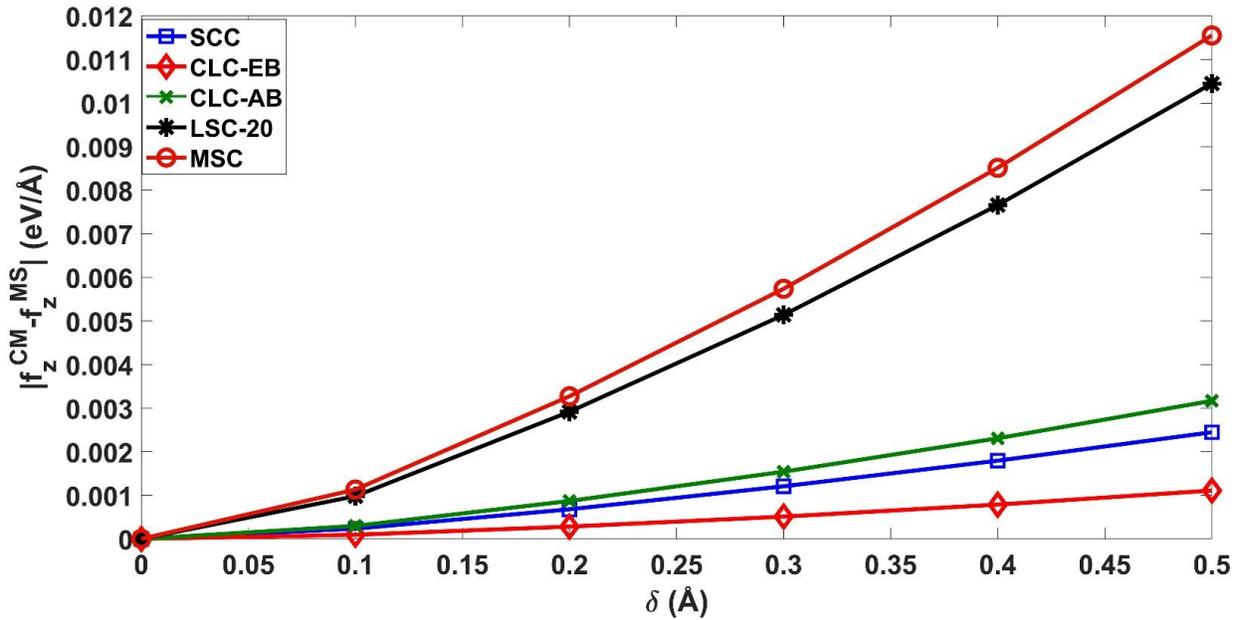

Fig. 21. Comparison of the error in the diamond tip reaction force between the SCC and the other weak compatibility couplings.

It is essential for the coupling methods to converge to the SCC method by decreasing the size of the elements near the interface. For some methods, this is evident. For example, if the size of the elements is sufficiently decreased until the number of nodes is equal to the number of atoms at the interface, DC (the method with the highest error) becomes SCC. Thus, we concentrate only on the convergence of the CLC-AB and CLC-EB schemes which might be less obvious. For this purpose, four new different meshes are generated for the continuum region illustrated in Fig. 8. Characteristics and details of the new meshes, as well as mesh types A and FR, are presented in Table 2. Furthermore, the number of interface atoms for all meshes equals 2441. All the mesh types but the FR mesh are arranged in alphabetical order from the coarsest mesh to the finest mesh ("A" is the coarsest and "E" is the finest).

Table 2
Mesh generation details.

| Mesh-type | ADBNIN* | Number of interface nodes | Average volume of interface elements (Å³) | Number of interface elements | Number of nodes | Number of elements |
|---|---|---|---|---|---|---|
| FR | 1 | 2441 | 5.4 | 13868 | 9520 | 47388 |
| A | 4.714 | 96 | 651.1 | 672 | 1415 | 6374 |
| B | 4.04 | 148 | 324.4 | 1103 | 1760 | 8220 |
| C | 2.828 | 321 | 102.5 | 2128 | 2544 | 12277 |
| D | 1.414 | 1241 | 13.1 | 7352 | 5785 | 28615 |
| E | 0.707 | 4881 | 1.7 | 27655 | 17405 | 86943 |

Note: * Average distance between nearest interface nodes scaled by the distance between nearest-neighbor atoms at the interface ($a_1\sqrt{2}/2$).

Again, the distribution of the interface nodes on all the boundary surfaces but the Z=30 $a_1$ surface is the same for all the mesh types, and therefore, the number of fixed atoms for all meshes equals 144. It is noteworthy that the interface elements in the type E mesh are even smaller than the interface elements in the type FR mesh, i.e., the number of interface nodes is more than the number of interface atoms, as can be seen in Fig. 22. The type E mesh is highly expensive, and it is simply generated to examine the flexibility of the CLC-AB scheme. As before, for all mesh types, loading is applied by moving semi-sphere 0.1 Å along the negative Z-axis for each increment to induce force on the substrate over five incremental loadings. The global displacement errors of the CLC-AB and CLC-EB schemes after the fifth incremental loading are plotted in Fig. 23 for the mesh types listed in Table 2. As can be seen in Fig. 23, by increasing the interface nodes ($N_I$) from the type A mesh to the type D mesh, the global displacement error for both schemes nearly decreases down to the error of the SCC method. It is worth noting that the error of the CLC-AB scheme equipped with the type D mesh is less than the SCC method. Continuing to decrease the size of the interface elements and therefore converting to the type E mesh increases the global displacement error. Increasing the error after the type D mesh can be considered the round-off error. The norm of the absolute displacements error of every atom for the CLC-AB scheme equipped with the meshes B, C, D, and E are illustrated in Fig. 24. As can be seen, while the interface elements getting smaller, the errors approach to the SCC method shown in Fig. 19(a). The errors for the CLC-EB are similar to Fig. 24.

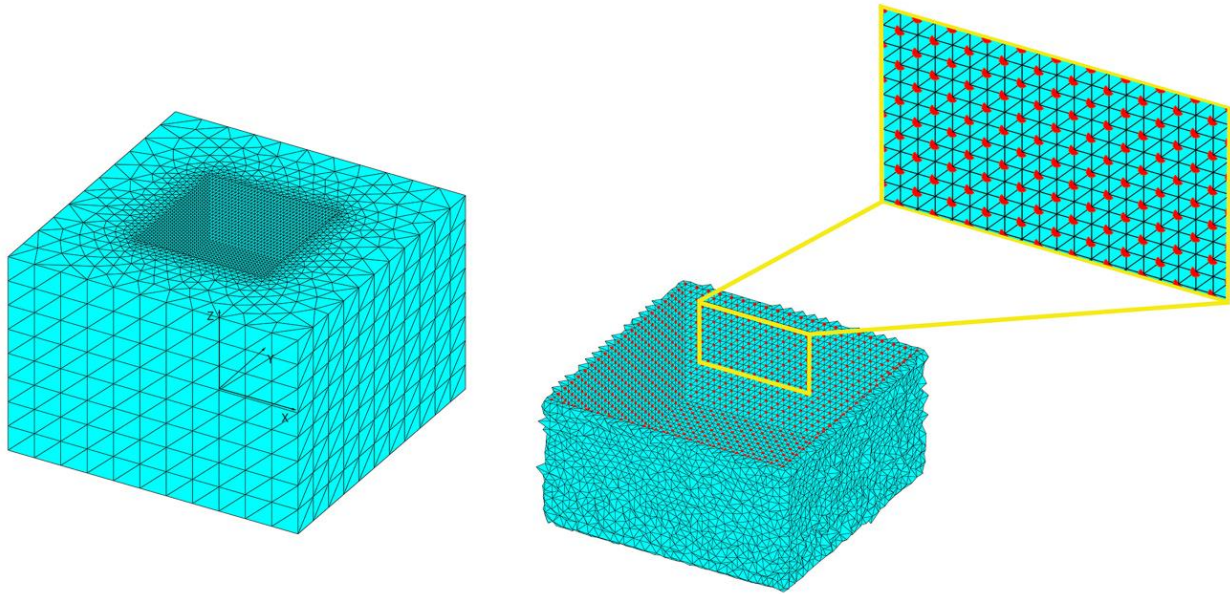

Fig. 22. (Left) Type E mesh and (right) its interface elements and atoms. The number of interface atoms (red spheres) is less than the number of nodes, as illustrated in the magnified view.

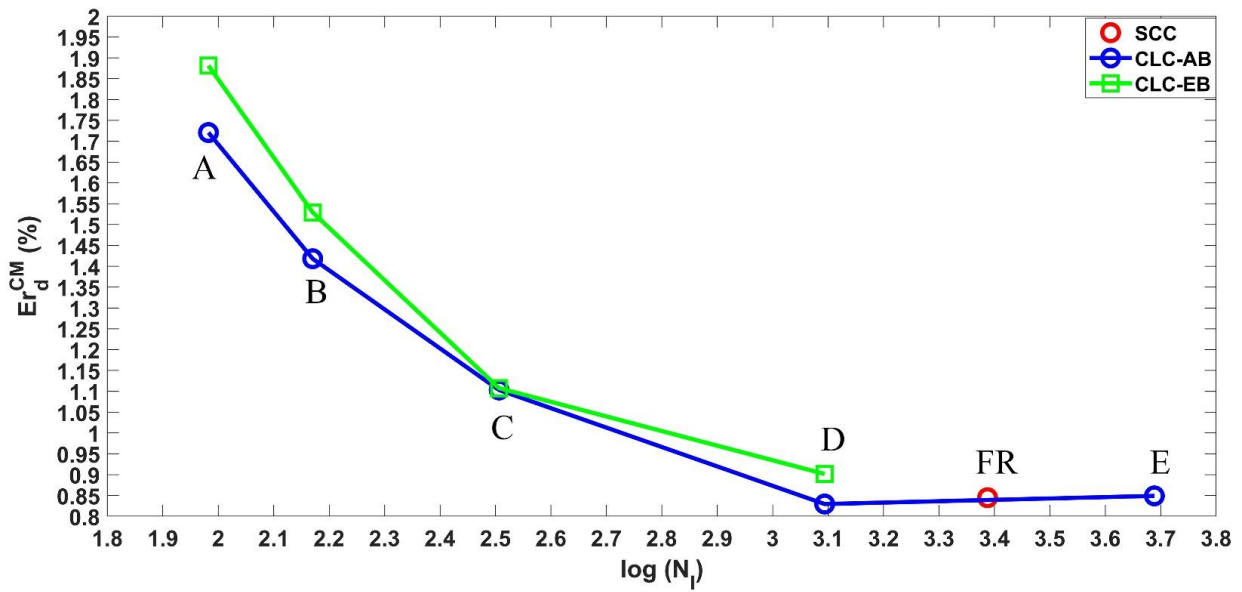

Fig. 23. Comparison of the global displacement errors between the CLC schemes equipped with the type A, B, C, D, and E meshes and strong compatibility coupling with the FR mesh.

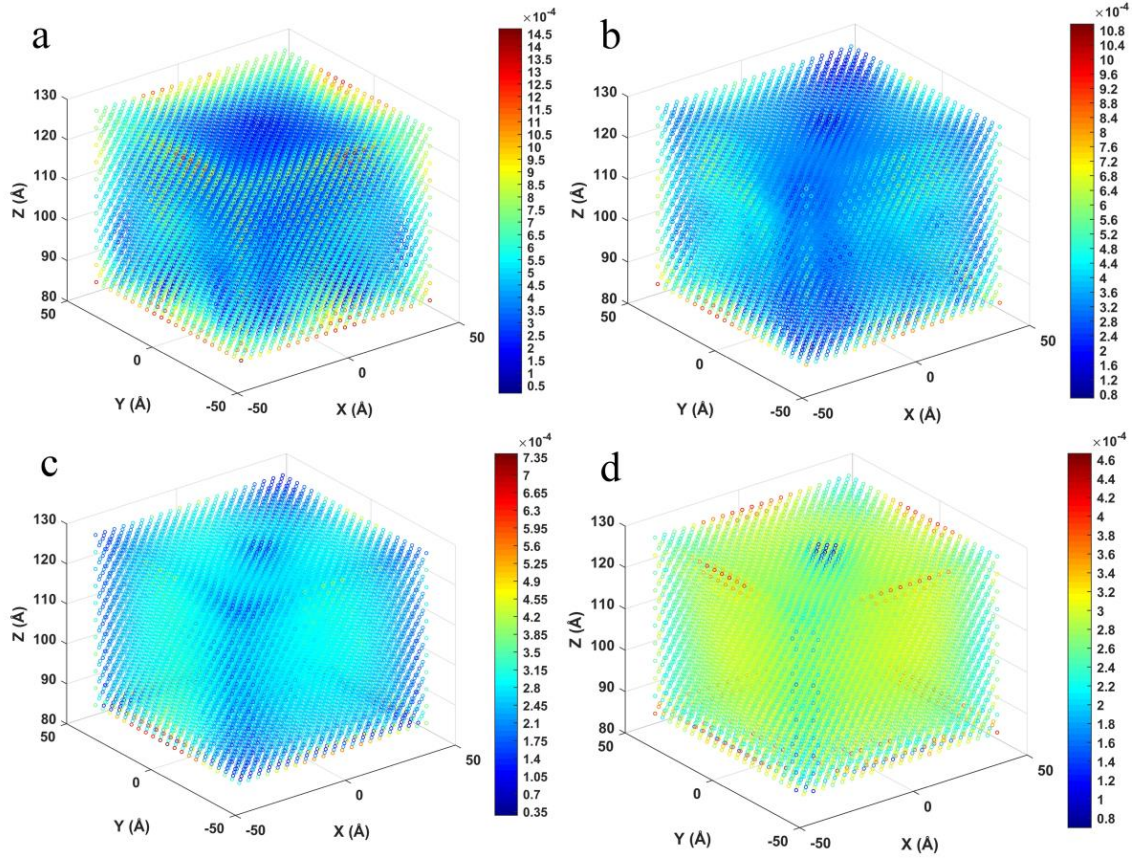

Fig. 24. Norm of the absolute displacements error of the substrate atoms for the CLC-AB scheme equipped with (a) type B mesh, (b) type C mesh, (c) type D mesh, and (d) type E mesh.

At this stage, we complete the discussion by reporting the average runtime over five increments for the SCC method and the CLC-AB scheme with different mesh types. The characteristics of the system we have used for the analysis are Intel(R) Core(TM) i5-1035G1 CPU processor with 8GB RAM, and the results are listed in Table 3. It is not pointless to mention that the runtime difference of the weak compatibility coupling methods is not substantial (and hence only the runtime for the CLC-AB scheme is reported) at least for our test problem. It is noteworthy to mention that the CLC-AB equipped with the type D mesh is superior to SCC due to its higher accuracy (Fig. (23)) and significantly lower computational cost (Table 3).

Table 3
Average runtime over five increments for the SCC and the CLC-AB methods.

| Coupling method | SCC | CLC-AB | | | | |
|---|---|---|---|---|---|---|
| Mesh type | FR | A | B | C | D | E |
| Hours | 4.87 | 0.64 | 0.73 | 1.25 | 2.77 | 9.28 |

Finally, we present the force-displacement curve, which is generally important in contact mechanics. To achieve a smoother curve, we reduce the loading increment to 0.05 Å (half of the previous simulations). The loading is imposed over 20 increments (1 Å movement of the semi-sphere along the negative Z-axis in total) with the same essential boundary conditions as before. Next, we choose the type C mesh for this simulation since, from Fig. 23 and Table 3, it is a reasonable choice for balancing the accuracy and computational cost. Given the type C mesh, both CLC-EB and CLC-AB schemes have resulted in nearly identical solutions, according to Fig. 23. Therefore, the CLC-EB scheme is only applied in this simulation and is regarded as the solution of the CLC method. The results for the reaction force in the Z-direction versus the displacements of the semi-sphere are plotted in Fig. 25. The reaction force of the semi-sphere, shown by blue line for the molecular statics (MS) and red crosses for the multiscale CLC method, is the summation of forces on the diamond atoms in the Z-direction. Another investigation of whether forces are transmitted correctly from the substrate's atomistic region to the substrate's continuum region or not is the examination of the reaction forces. Since the problem is static, the reaction force of the semi-sphere must be equal to the reaction force of the substrate theoretically. In Fig. 25, green squares represent the sum of reaction forces in the Z-direction on all nodes of the substrate boundary on the Z=0 plane. As can be seen, green squares are coincident with the red crosses, which means the reaction force of the semi-sphere is equal to the reaction force of the substrate boundary. The total potential energy curve for the described problem is plotted in Fig. 26. It is important to note that in spite of avoiding double counting of energy at the interface, the total potential energy in the fully atomistic model is not equal to the multiscale model since the energies between the boundary atoms (atoms on the $X = -22.5\ a_1, X = 22.5\ a_1, Y = -22.5\ a_1,\ Y = 22.5\ a_1, Z = 0, and\ Z = 30\ a_1$ surfaces) are fully computed in the fully atomistic model, but only part of the energies are computed in the multiscale model. For this reason, the difference of the energy between the fully atomistic model and the multiscale model in the initial state is added to the multiscale model as a constant energy. Positions of the substrate atoms on the uppermost layer (atoms on the $Z = 30\ a_1$ surface) for the multiscale CLC model and fully atomistic model before loading and after the 20th loading increment are shown in Figs. 27(a) and (b), respectively. As can be seen in Fig. 27(b), atoms on the Z=30 surface are perfectly matched for the two models after loading. Displacement field of the multiscale model in the Z-direction before and after loading are shown in the Figs. 27(c) and (d).

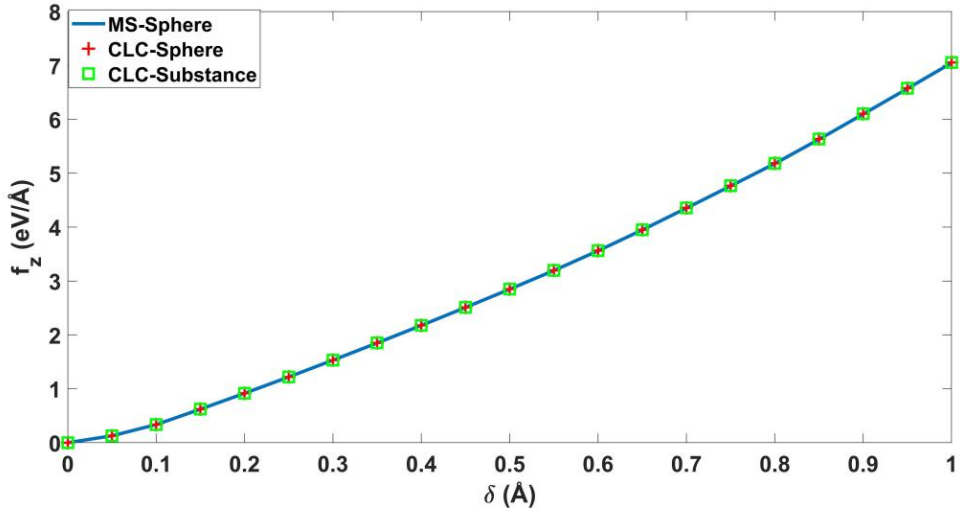

Fig. 25. Comparison of the reaction force between MS and CLC. MS-Sphere, CLC-Sphere, and CLC-substance stand for reaction force along Z-axis for sphere calculated by MS, for sphere calculated by CLC, and for substance calculated by CLC, respectively.

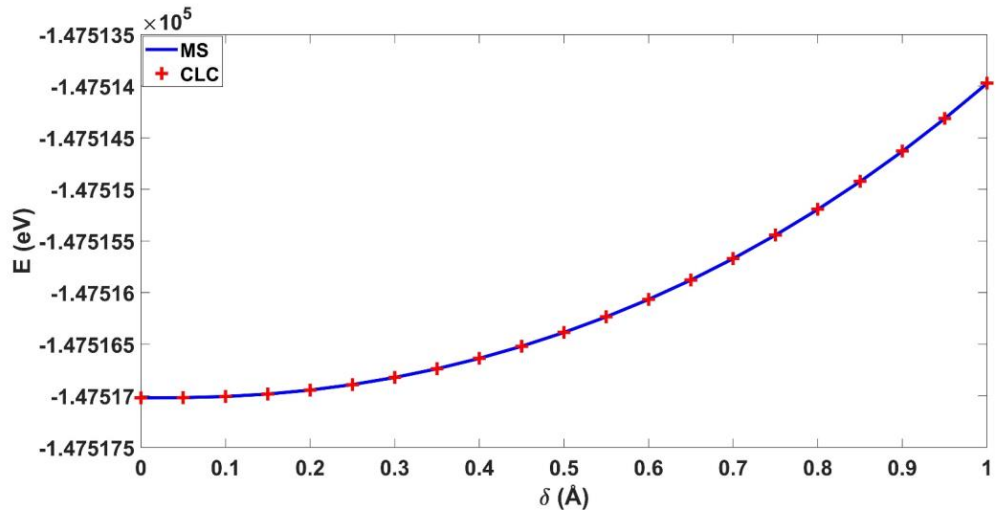

Fig. 26. Comparison of the total energy between MS and CLC.

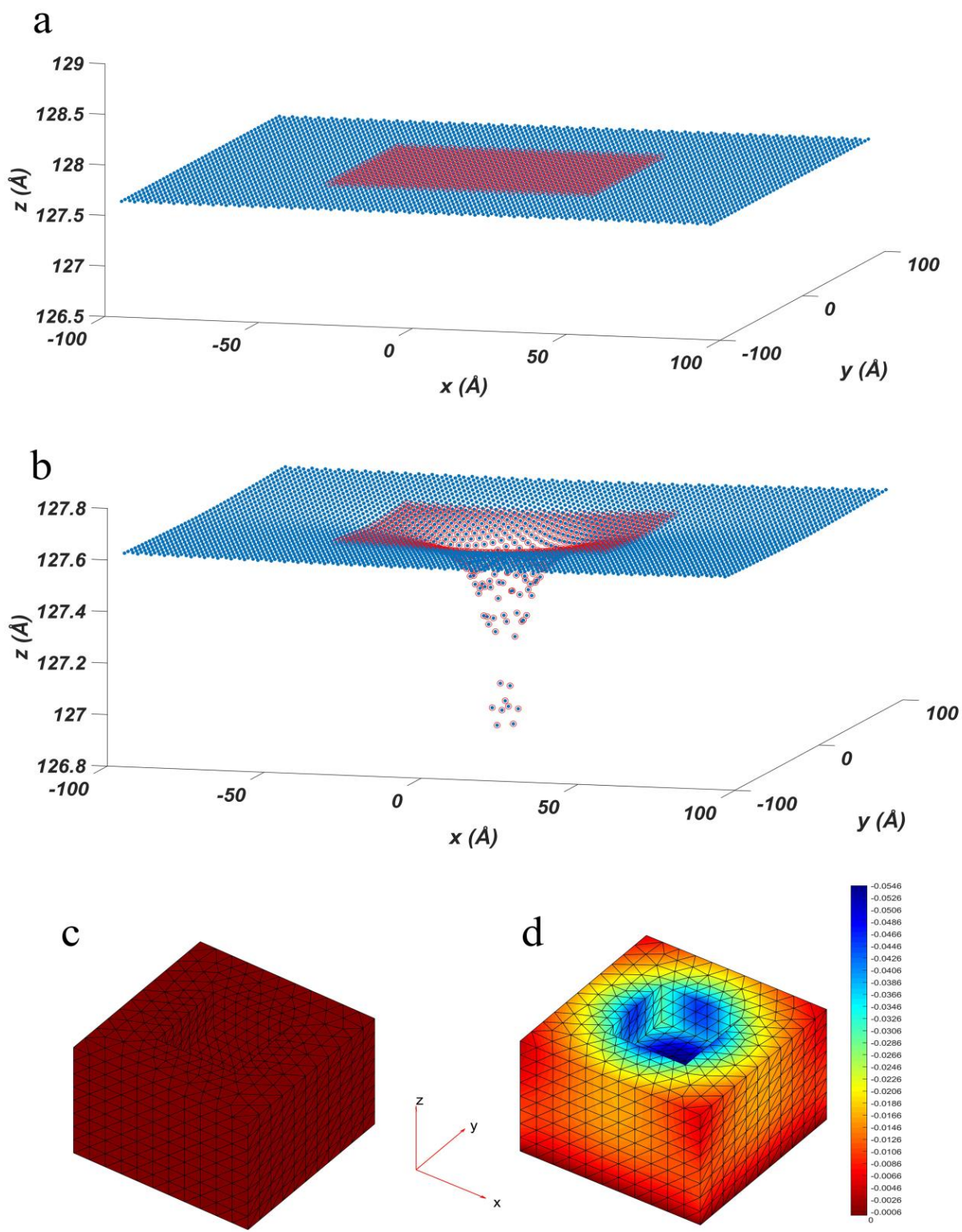

Fig. 27. Substrate before and after loading. Red circles are atoms in the multiscale model, and blue spheres are atoms in the fully atomistic model. (a) Atoms on the $Z = 30\ a_1$ surface in the initial state. (b) Atoms on the $Z = 30\ a_1$ surface after 20th incremental loading. The displacement field (in angstrom) of the finite elements part in the Z-direction: (c) in the initial state and (d) after 20th incremental loading.

# 7. Consistent linear coupling method as a way to make consistency

In previous sections, coupling atoms with nodes at the interface was discussed. The total energy was introduced in Eq. (82) and was minimized to obtain the displacements in the nanoscale contact problem. However, $E_{consistancy}$ has been referred to this section, and it was considered equal to zero there. In this section, we present how the consistent linear coupling (CLC) method is able to address the $E_{consistancy}$.

## 7.1. Ghost force problem and a theoretical treatment

The standard continuum model is local by definition. On the other hand, atomistic models with empirical potentials require a finite range of interaction to predict material behavior accurately and hence are considered nonlocal generally. As a result, an inconsistency was observed by coupling the two models since they are different in nature in the sense that one is nonlocal and the other is local [20]. The inconsistency causes the "ghost forces," which are even present in the equilibrium configuration. Several methods have been proposed to remove the ghost forces. For example, the ghost forces can be removed by adding dead loads [20]. Although the method has the advantage of simple implementation, the forces are nonconservative. In more fundamental work, the special atoms are designated between nonlocal and local regions (so-called "quasi-nonlocal" atoms) to interact with local regions locally and with nonlocal regions nonlocally [45]. The forces in this approach remain conservative to a certain range of interaction. For instance, in a one-dimensional atomic chain, it is limited to the second-nearest-neighbor interactions. More sophisticated methods were developed later to break the limit, e.g., the reconstruction scheme [46], the method of combining the exact and the continuum contributions [47], and the blended ghost force correction method [48, 49]. Here, we offer a closed-form solution for $E_{consistancy}$ to remove the ghost forces for an arbitrary range of interaction by applying the CLC method. Therefore, the forces are conservative in this approach.

To express the problem of the ghost force, we describe the conventional multiscale method as follows. Consider a one-dimensional atomic chain in the equilibrium configuration. The atomic chain is split into the atomistic and continuum parts, and the continuum part is discretized by the finite elements, as shown in Fig. 28. Then, consider a pair potential limited to the second-nearest-neighbor interaction and focus on the part of the energy involving atom - 1. The energy between atoms -1 and -3 is treated atomistically. In the case of energy between atom -1 and node 1, half of this energy is computed by the energy of the element $el_1$ and the other half is computed atomistically. Therefore, the sum of these two terms gives the true energy between atom -1 and node 1 in the reference configuration. Although it still seems to be no problem since the initial energy is computed correctly near the interface, the difficulty appears by calculating force. Indeed, the force associated with the energy between atoms -1 and node 1 is nonlocal, but the force associated with the energy of the element $el_1$ is local, and rather than exerting force on atom -1, it exerts a force on node 0. This introduces an unbalanced force (ghost force) on atom -1 in the equilibrium configuration. Consequently, the conventional multiscale method (all energy-based multiscale methods [8]) suffers from the ghost forces, and some additional efforts are required to remove them.

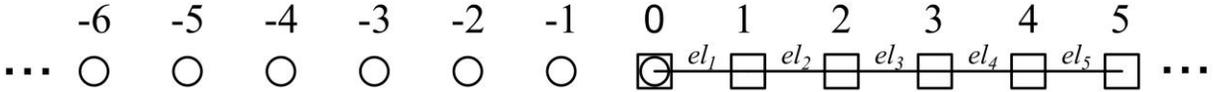

Fig. 28. Conventional multiscale model. Circles are atoms, and squares are nodes. Atom 0 is the interface atom.

What we seek to define in this section is the total energy (Eq. (82)) in order to remove the ghost forces. Herein, we only concentrate on the pair potentials (additive pair-wise interactions). The energy of the coarse scale, $\widetilde{\Pi}$, is given by Eq. (28). The reduced atomistic energy, $\hat{E}_{atomistic}$, is computed by ignoring the continuum part, i.e., only the energy between atoms is considered. For instance, in contrast to the conventional multiscale method, energy between atom –1 and node 1 in Fig. 28 is not calculated by $\hat{E}_{atomistic}$. Therefore, the only undefined term is $E_{consistency}$. At this point, half of the energies between nodes and atoms near the interface are calculated by the continuum part ($\widetilde{\Pi}$) and the other half is not considered by the definition of $\hat{E}_{atomistic}$. To compensate for the energy deficiency, we first add new elements and nodes to the system, as shown in Fig. 29. The positions of the added nodes are arranged in a specific pattern to eliminate the ghost forces, as will be revealed afterward. Depending on the maximum range of interatomic potential, say $N$, elements and nodes are added from the second to $N$th row according to Fig. 29 pattern corresponding to the second to $N$th-nearest neighbors. In Fig. 29, the reference configuration is considered to be the equilibrium configuration, and the elements associated with the $n$th row are only responsible for computing the energy between $n$th neighbors. For instance, consider an interatomic potential energy with a maximum range of third-nearest-neighbor interaction ($N = 3$). The energies of the elements in the second and third rows are computed by considering only the interactions between the second- and third-nearest neighbors of underlying atoms, respectively, as illustrated in Fig. 30.

Recall that the energy of the elements based on the underlying atoms with representative atom $i^*$ is given by Eq. (19). Next, we number the added nodes and elements from 1 to $M$ and 1 to $M - 1$ (from left to right), respectively, for every row. The energy for consistency up to $N$th-nearest-neighbor interactions is defined as

$$E_{consistency} := \sum_{n=2}^{N} \sum_{m=1}^{M-1} E_n^{el'_m} - \hat{E}^{el_1}, \tag{94}$$

where $n$ is the row number, shown in Fig. 29, $m$ is the element number, $E_n^{el'_m}$ is the energy calculated between $n$th neighbors associated with the added element m, and $M$ is the number of nodes, which depends on the $n$th row and is given by

$$M = \begin{cases} \dfrac{n+2}{2}, & n \text{ even} \\ \dfrac{n+1}{2}, & n \text{ odd} \end{cases} \tag{95}$$

In Eq. (94), participating exclusively the first term without the second term overestimates the total energy. Therefore, the second term was introduced to reduce the energy to the actual value. The second term in Eq. (94), $\hat{E}^{el_1}$, is the energy of the element $el_1$ considering only half of the interactions of the even neighbors. The underlying atoms associated with the element $el_1$ for calculating $\hat{E}^{el_1}$ is shown in Fig. 31. It should be noted that for the special case when $N = 1$ (nearest-neighbor interatomic potential), $E_{consistency} := 0$, as was applied for the nanoscale contact problem.

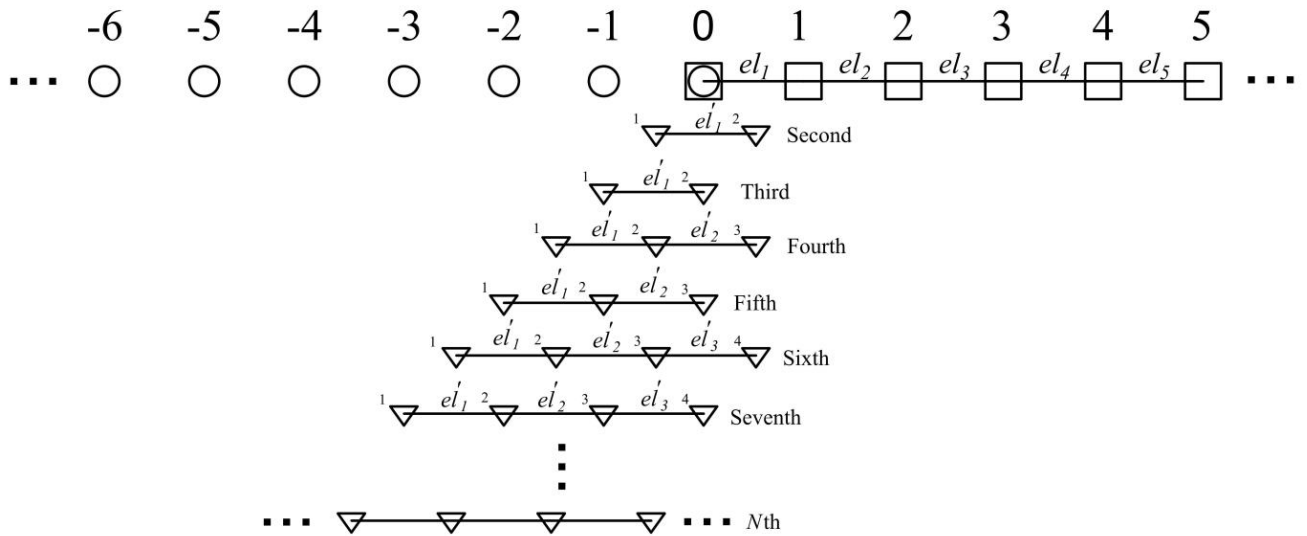

Fig. 29. The CLC multiscale model. Circles are atoms, squares are real nodes, and triangles are the added nodes arranged in specific pattern. Note that all rows lie on the first row (on the principal multiscale model), and they are illustrated separately for more clarity.

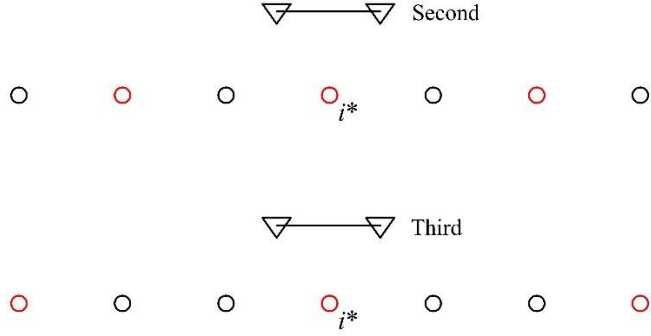

Fig. 30. Underlying atoms associated with the second- and third-row elements. In computing the energy of the elements, red atoms are only considered.

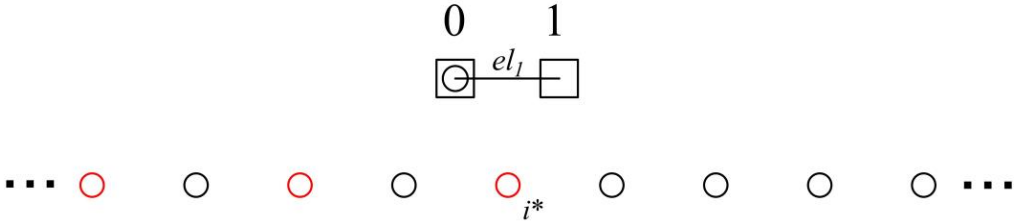

Fig. 31. Underlying atoms associated with the element $el_1$ in terms of computing $\hat{E}^{el_1}$, in which only red atoms are considered.

Thus far, we have completed the definition of the total energy (Eq. (82)) by means of adding new nodes and elements to the system. However, adding new nodes increases the degrees of freedom of the system. The added nodes need to be constrained in terms of atoms and real nodes not only to reduce the degrees of freedom to the actual value but also to eliminate the ghost forces. For defining constraints of the added nodes, we return to the definition of the consistent linear coupling (CLC) method introduced by Eq. (35). Similar to the previous discussion but with a subtle difference, the coefficients $c_i^\alpha$ are determined so that the forces on atoms near the interface be the same as the fully atomistic model (instead of the strong compatibility model). By doing so, the displacements of the added nodes can be written as

$$\boldsymbol{q}_n^m = \frac{1}{n}\sum_{i=0}^{n-1} \boldsymbol{u}_{0-i+m-1}, \quad n = 2, \ldots, N \text{ and } m = 1, 2, \ldots, M-1 \tag{96}$$

$$\boldsymbol{q}_n^M = \begin{cases} \dfrac{\boldsymbol{u}_0 + \boldsymbol{u}_1}{2}, & n \text{ even} \\ \boldsymbol{u}_0, & n \text{ odd} \end{cases} \tag{97}$$

where $\boldsymbol{q}_n^m$ is the displacement of the added node placed in the $n$th row and $m$th column (from left to right) of the pattern, $M$ is given by Eq. (95), and $\boldsymbol{u}_{0-i+m-1}$ is the displacement of the atom or real node depending on the subscript of $\boldsymbol{u}$ attached to every atom or real node according to Fig. 29. By considering the consistency energy definition in Eq. (94), the pattern of the added nodes in Fig. 29, and the constraint Eqs. (96) and (97) we claim that the geometry, force, and energy are consistent in equilibrium configuration. See Appendix for the proofs.

## 7.2. Validation

The introduced multiscale methods, conventional and CLC, are compared with each other by considering the fact that the solution of the fully atomistic model is the exact solution. To examine how the CLC method is effective, we relax the total energy of a one-dimensional atomic chain consisting of 21 equally-spaced atoms. We describe the test problem as follows. The fully atomistic and multiscale models are shown in Figs. 32 and 33, respectively. The same aluminum interatomic potential described in Eq. (10) is adopted for all models and methods. However, the cutoff radius is replaced by $6\sigma$ to increase the maximum range of interaction to include up to the fifth-nearest neighbor. The boundary atoms 1 and 21 in Fig. 32 are set free and fixed, respectively, and no external loading is applied. In the case of the fully atomistic model, 4 more atoms are created next to atom 21 (unnumbered atoms in Fig. 32) and held fixed to eliminate the surface effect on the atoms near atom 21. For the multiscale model, the 21 atoms are partitioned into 10 regular atoms, 1 interface atom (atom number 11), and 10 nodes, as shown in Fig. 33. Note that the degrees of freedom are equal for both models and methods. The boundary conditions correspond to the fully atomistic model. Thus, atom 1 is set as a free boundary condition, and node 21 is held fixed during relaxation. For both conventional and CLC multiscale methods, the initial energy is equal to -8.10764 eV, whereas the initial energy for the fully atomistic model is equal to -8.11531 eV. The difference between the initial energy of the multiscale models with the fully atomistic model initiates from the full contribution of 4 more added atoms to the energy in the fully atomistic model. The absolute displacement error of atoms and nodes for the conventional and CLC multiscale methods are plotted in Figs. 34 and 35, respectively. The global displacement and energy errors based on all atoms are calculated via Eqs. (92) and (93) and presented in Table 4. Results demonstrate that the ghost forces can cause high errors near the interface for the conventional multiscale method. On the contrary, the CLC method can eliminate the ghost forces and produces an extremely accurate solution.

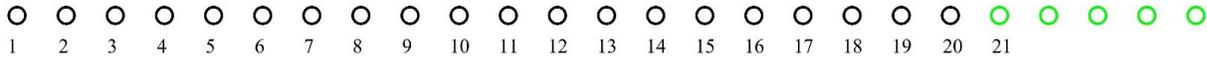

Fig. 32. Fully atomistic model. The green atoms (number 21 and unnumbered atoms) are held fixed during relaxation.

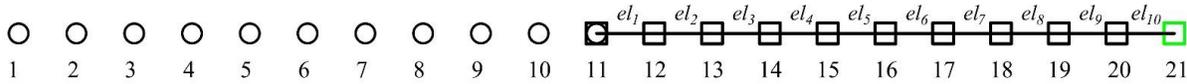

Fig. 33. Multiscale model. The green node (number 21) is held fixed during relaxation.

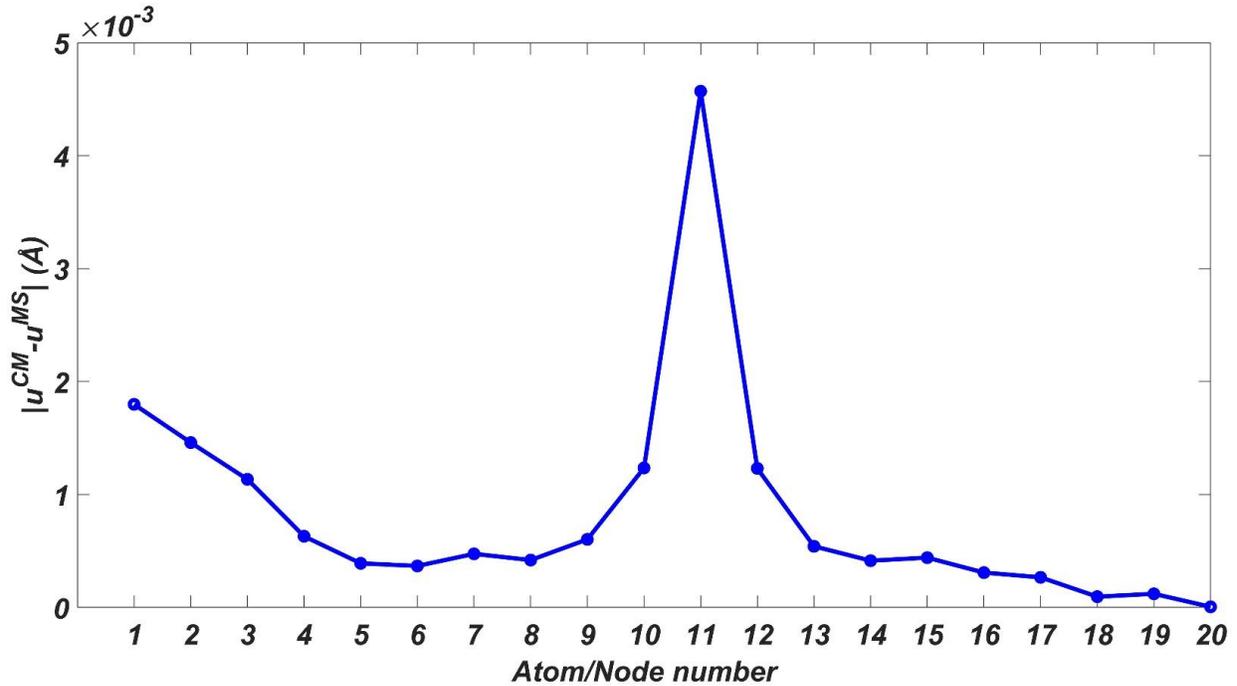

Fig. 34. Absolute displacement error of atoms/nodes in the conventional multiscale method. CM and MS stand for conventional multiscale and molecular statics methods, respectively.

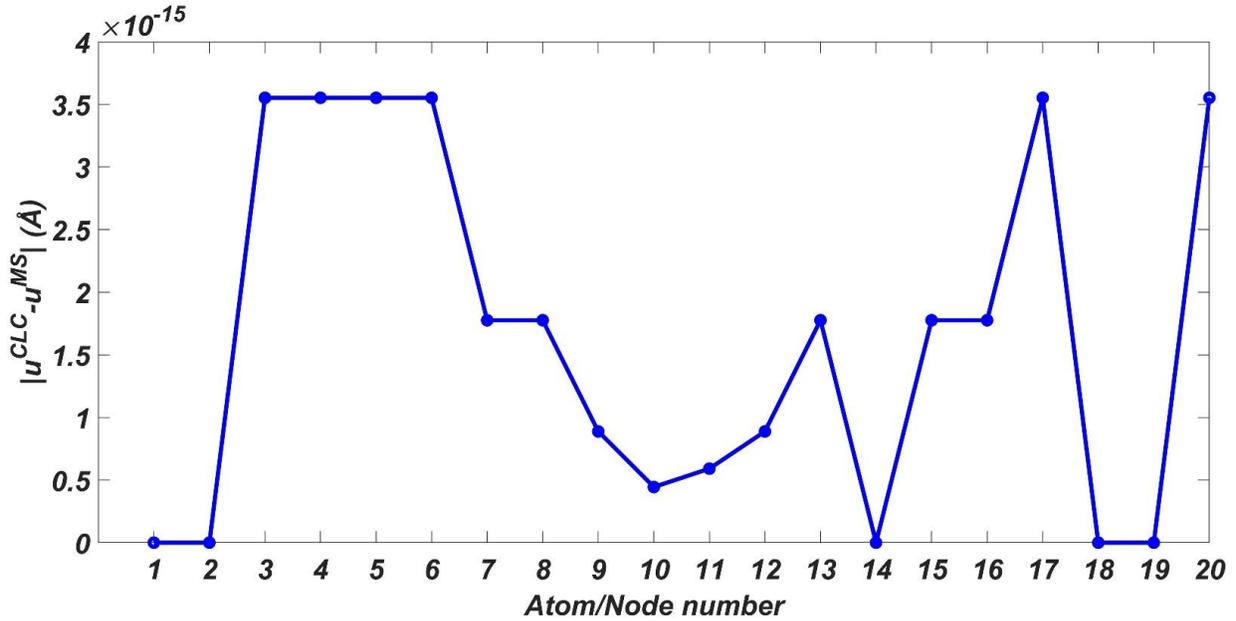

Fig. 35. Absolute displacement error of atoms/nodes in the CLC multiscale method. CLC and MS stand for consistent linear coupling and molecular statics methods, respectively.

Table 4: Global displacement and energy errors in the conventional and CLC multiscale methods.

| Multiscale method | Global displacement error (%) | Energy error (%) |
|---|---|---|
| Conventional | 124 | 26.47 |
| CLC | $2.1 \times 10^{-10}$ | $2.9 \times 10^{-8}$ |

## 8. Discussion

Among the methods that are based on the surface approximation approach, the LS-$n_{at}$ and DC method can show the lowest and highest errors, respectively. The poor accuracy of the DC method stems from the fact that many interface atoms do not contribute explicitly to constraining the interface nodes in this method. On the other hand, the accuracy of the LS-$n_{at}$ method highly depends on the set $n_c$. In this paper, we have considered identical $n_{at}$ for all nodes, and we have found that when the $n_{at}$ equals 40, the global displacement error was minimum in terms of solving the nanoscale contact problem. However, finding such a value beforehand, or more satisfyingly determining $n_{at}$ specifically for each node remains as future work.

Despite the fact that both SCC and MSC preserve the continuity along the interface, the results have revealed that the SCC is considerably more accurate. This is because, in the SCC, the inner surface belongs to the more accurate model ($\Gamma_1$) governs the inner surface belongs to the less accurate model ($\Gamma_2$), i.e., $\Gamma_1$ is the principal surface. On the contrary, in the MSC, $\Gamma_2$ is the principal surface. Given the fact that the MSC is less accurate than the other coupling methods that set $\Gamma_1$ as the principal surface, such as SCC, CLC-based schemes, and LS-40, we deduce that in coupling models with individually different accuracy, the surface belonging to a more accurate model should essentially govern the surface belonging to a less accurate model.

Comparing CLC-AB with CLC-EB: Although we have developed the CLC method in the ideal case, we have applied it to the challenging condition where the interface had sharp corners and edges. As was expected by theory, corners and edges have shown maximum errors for both CLC schemes. However, with a nearly regular distribution of nodes (which is common in any mesh generation software) on the interface, acceptable accuracy can be obtained even with the large interface elements. The CLC-AB scheme is less sensitive to deviation from the ideal case. On the other hand, the CLC-AB scheme is supposed to be more accurate for the condition near the ideal case with a relatively high varying deformation gradient at the interface. Therefore, the CLC-AB scheme is a reasonable choice for a general interface, whereas the CLC-EB is recommended for a sufficiently flat interface. The CLC-AB scheme is more flexible as we could apply it to the type E mesh, in which the number of nodes is more than the number of atoms at the interface. Finally, depending on which user wants to weigh more, accuracy or reduction of computational cost, we would recommend the distance between the nodes on the interface for the CLC-based schemes should be something between the distance of the nearest-neighbor atoms in a crystal and three times of that.

Comparing CLC-AB with SCC: To the best of our knowledge, multiscale methods that employ the SCC, such as QC, CLS, CADD, and FEAt are the most accurate coupling methods, as the multiple studies confirmed [33, 34]. According to our numerical results, the CLC-AB surpass the SCC as the global displacement error for the SCC and the CLC-AB were 0.845 and 0.830 percent, respectively, and the average runtime for the SCC and the CLC-AB were 4.87 and 2.77 hours, respectively. It is true that we established the CLC method based on the SCC, and therefore SCC must be more accurate. However, we can name two reasons why the CLC-AB has shown slightly higher accuracy. First, in the SCC, the size of the interface elements is minimal due to the constraining of the interface nodes, and therefore the round-off error can be more than CLC (for the reasonable size of the interface elements, e.g., type D mesh in the nanoscale contact problem). Second, the mesh quality in the CLC-AB scheme can be higher due to the non-restricted mesh generation. Indeed, such superiorities can cover the error of deviation from the ideal case in the CLC-AB scheme.

## 9. Conclusions

The main aim of this paper was to investigate the possibility of early mesh coarsening at the atomistic-continuum interface in order to reduce the computational cost substantially with slightly (or even without) sacrificing accuracy. To this end, the conditions ensuring the continuity between models and relaxing the strong compatibility have been discussed. Accordingly, besides almost known weak coupling methods, such as direct coupling, least-squares coupling, and master-slave coupling, we have presented a new weak compatibility coupling method called consistent linear coupling (CLC) and its schemes, CLC-AB and CLC-EB, for the purpose of early mesh coarsening. Moreover, a new approach to avoid double counting of energy at the interface has been proposed. The weak compatibility couplings and the most accurate coupling method, the strong compatibility coupling (SCC), have been compared in solving the nanoscale contact problem. The CLC-based schemes have demonstrated the best accuracy among the introduced methods. In particular, the CLC-AB scheme equipped with the type D mesh compared with the SCC has indicated significantly less computational cost and could remarkably preserve accuracy. The features of the CLC can be enumerated as follows:

- Easy implementation as no specific mesh generation algorithm is required, unlike the SCC.
- High-quality mesh is achievable due to non-restricted mesh generation.
- Flexibility: Wide range of sizes of the interface element can be selected, enabling optimizing the computational cost and accuracy.
- Convergence: By reducing the size of the interface elements, the accuracy is increased.
- Robustness: Despite deviation from the ideal case, it shows acceptable accuracy.

Finally, we have applied the CLC concept to produce matching force conditions on the atoms near the interface in order to eliminate ghost forces in the multiscale model for the one-dimensional atomic chain. After relaxation, the result was in perfect agreement with the fully atomistic model. Unfortunately, generalization to higher dimensions is not straightforward and remains as future work.

# Appendix

Here, we demonstrate that the definition of the consistency energy (Eq. (94)) and the constraints in Eqs. (96) and (97) in Section 4 lead to the geometrical, force, and energy consistency at and near the atomistic-continuum interface. Note that the label of the atoms, real nodes, and added nodes is identical to the given labels in Fig. (29).

## A.1. Geometrical consistency

The added-constrained nodes (triangles in Fig. 29) in the reference configuration have been arranged in order to be the average of the positions of their corresponding atoms and real nodes. Therefore, one can easily verify that Eqs. (96) and (97) are satisfied in the reference configuration, i.e.,

$$Y_n^m = \frac{1}{n} \sum_{i=0}^{n-1} Z_{0-i+m-1}, \quad n = 2, \dots, N \text{ and } m = 1, 2, \dots, M-1 \tag{A.1}$$

$$Y_n^M = \begin{cases} \dfrac{Z_0 + Z_1}{2}, & n \text{ even} \\ Z_0, & n \text{ odd} \end{cases} \tag{A.2}$$

where $Y_n^m$ is the position of the added node in the reference configuration, and $Z_{0-i+m-1}$ depends on its subscript index is the position of the atom or real node in the reference configuration.

## A.2. Force consistency

As the forces on the atoms are balanced in the equilibrium state in the fully atomistic model, the force consistency is valid when the forces on the atoms and nodes are balanced in the equilibrium state in the multiscale model. Let U be the parts of the interatomic potential energy that is responsible for forces on atom $\alpha$, $\phi_n^\alpha$ be the energy between an individual atom/node $\alpha$ and its nth-nearest neighbors, and assume that $B^A$ is the set of atoms including regular and interface ($\alpha = 0$ in Fig. 29) atoms. In the following conditions, the balance of forces on atom $\alpha$ interacting with its $n$th neighbors is examined. The conditions are based on the position of atom/node $\alpha$ and the value of $n$, and hence they are subdivided into four main groups as below

(1) If $\alpha + n \in B^A$ and $\alpha - n \in B^A$

In this condition, the energy associated with the atom $\alpha$ is written as

$$\phi_n^\alpha = E_n^{\alpha,\alpha-n} + E_n^{\alpha,\alpha+n} \tag{A.3}$$

where the energy term $E_n^{a,b}$ is the energy between atoms a and b. The energy $E_n^{a,b}$ is the function of positions of atoms a and b in the current configuration. Therefore we write $E_n^{a,b} = \varphi(r^{a,b})$ with $|b - a| = n$, where $r^{a,b}$ is the distance between atoms a and b. In the first condition, U is equal to $\phi_n^\alpha$ therefore the force on atom $\alpha$ is equal to

$$\frac{\partial U}{\partial r^\alpha} = \frac{\partial \phi_n^\alpha}{\partial r^\alpha} = \frac{\partial E_n^{\alpha,\alpha-n}}{\partial r^\alpha} + \frac{\partial E_n^{\alpha,\alpha+n}}{\partial r^\alpha}, \tag{A.4}$$

and in the equilibrium state, we have

$$\frac{\partial U}{\partial r^\alpha} = \frac{\partial \varphi(r^{\alpha,\alpha-n})}{\partial r^\alpha} + \frac{\partial \varphi(r^{\alpha,\alpha+n})}{\partial r^\alpha} = 0. \tag{A.5}$$

(2) If $\alpha + n \notin B^A, \alpha - n \in B^A, \alpha \in B^A$ and $\alpha \neq 0$

In this condition, the energy associated with the atom $\alpha$ is written as

$$\phi_n^\alpha = E_n^{\alpha,\alpha-n} \tag{A.6}$$

In contrast to the previous condition, U is not equal to $\phi_n^\alpha$. And the force on atom $\alpha$ can be calculated as

$$\frac{\partial U}{\partial \boldsymbol{r}^\alpha} = \frac{\partial \phi_n^\alpha}{\partial \boldsymbol{r}^\alpha} + \sum_{i=1}^{M-1} \sum_m \frac{\partial E_n^{el'_i}}{\partial \boldsymbol{q}_n^m} \frac{\partial \boldsymbol{q}_n^m}{\partial \boldsymbol{r}^\alpha} \tag{A.7}$$

where $\boldsymbol{q}_n^m$ is the displacement of the added node (see Eq. (96) and (97)). Now let $\boldsymbol{p}_n^m$ and $\boldsymbol{p}_n^M$ be the position of the added node, i.e., $\boldsymbol{p}_n^m = \boldsymbol{Y}_n^m + \boldsymbol{q}_n^m$ and $\boldsymbol{p}_n^M = \boldsymbol{Y}_n^M + \boldsymbol{q}_n^M$. The second term in the right hand of Eq. (A7) can be simplified to

$$\sum_{i=1}^{M-1} \sum_m \frac{\partial E_n^{el'_i}}{\partial \boldsymbol{q}_n^m} \frac{\partial \boldsymbol{q}_n^m}{\partial \boldsymbol{r}^\alpha} = \frac{\partial \varphi(np_n^{1,2})}{\partial \boldsymbol{p}_n^1} \times \frac{\partial \boldsymbol{p}_n^1}{\partial \boldsymbol{r}^\alpha} \tag{A.8}$$

where $\boldsymbol{p}_n^1$ and $\boldsymbol{p}_n^2$ are the positions of the first and second added nodes (from left to right) associated with the $n$th row, respectively, and $p_n^{1,2}$ is the distance between atoms 1 and 2. Thus, all the forces on atom $\alpha$ can be calculated as

$$\frac{\partial U}{\partial \boldsymbol{r}^\alpha} = \frac{\partial E^{\alpha,\alpha-n}}{\partial \boldsymbol{r}^\alpha} + \frac{\partial \varphi(np_n^{1,2})}{\partial \boldsymbol{p}_n^1} \times \frac{\partial \boldsymbol{p}_n^1}{\partial \boldsymbol{r}^\alpha} = \frac{\partial \varphi(r^{\alpha,\alpha-n})}{\partial \boldsymbol{r}^\alpha} + \frac{1}{n}\frac{\partial \varphi(np_n^{1,2})}{\partial \boldsymbol{p}_n^1} = 0. \tag{A.9}$$

(3) If $\alpha \notin B^A$ and $\alpha \neq 1$

The region after node 1 (nodes 2,3, …) is kept untouched and can be treated as usual by the Cauchy-Born rule approximation given by Eqs. (20) and (28). What might still be ambiguous are the forces on the interface atom 0 and node 1.

(4) If $\alpha = 0$ or $\alpha = 1$

We investigate both separately for two states, first when $n$ is even, and second when $n$ is odd.

    (i)    For $\alpha = 1$ and even $n$, considering the second term in Eq. (94), we have

$$\phi_n^1 = \frac{1}{2}E_n^{el_1} + E_n^{el_2}. \tag{A.10}$$

The force on atom 1 can be calculated as

$$\frac{\partial U}{\partial \boldsymbol{r}^1} = \frac{\partial \phi_n^1}{\partial \boldsymbol{r}^1} + \frac{\partial E_n^{el'_M}}{\partial \boldsymbol{r}^1} = \frac{1}{2}\frac{\partial E_n^{el_1}}{\partial \boldsymbol{r}^1} + \frac{\partial E_n^{el_2}}{\partial \boldsymbol{r}^1} + \frac{\partial E_n^{el'_M}}{\partial \boldsymbol{r}^1}. \tag{A.11}$$

Considering Eq. (97), we have $\partial \boldsymbol{p}_n^M / \partial \boldsymbol{r}^1 = 1/2$. Thus

$$\frac{\partial U}{\partial \boldsymbol{r}^1} = \frac{1}{2}\frac{\partial \varphi(nr^{1,0})}{\partial \boldsymbol{r}^1} + \frac{\partial \varphi(nr^{1,2})}{\partial \boldsymbol{r}^1} + \frac{\partial \varphi(np_n^{M,M-1})}{\partial \boldsymbol{p}_n^M}\frac{\partial \boldsymbol{p}_n^M}{\partial \boldsymbol{r}^1} = 0. \tag{A.12}$$

    (ii)    For $\alpha = 0$ and even $n$, considering the second term in Eq. (94), we have

$$\phi_n^0 = \frac{1}{2}E_n^{el_1} + E_n^{0,-n} \tag{A.13}$$

Then, considering Eq. (97), we have $\partial \boldsymbol{p}_n^M / \partial \boldsymbol{r}^0 = 1/2$ and $\partial \boldsymbol{p}_n^1 / \partial \boldsymbol{r}^0 = 1/n$. Thus

$$\frac{\partial U}{\partial \boldsymbol{r}^0} = \frac{\partial \phi_n^0}{\partial \boldsymbol{r}^0} + \frac{\partial \varphi(np_n^{M,M-1})}{\partial \boldsymbol{p}_n^M}\frac{\partial \boldsymbol{p}_n^M}{\partial \boldsymbol{r}^0} + \frac{\partial \varphi(np_n^{1,2})}{\partial \boldsymbol{p}_n^1}\frac{\partial \boldsymbol{p}_n^1}{\partial \boldsymbol{r}^0} \tag{A.14}$$

$$= \frac{1}{2}\frac{\partial \varphi(nr^{1,0})}{\partial \boldsymbol{r}^0} + \frac{\partial \varphi(r^{0,-n})}{\partial \boldsymbol{r}^0} + \frac{\partial \varphi(np_n^{M,M-1})}{\partial \boldsymbol{p}_n^M}\frac{\partial \boldsymbol{p}_n^M}{\partial \boldsymbol{r}^0} + \frac{\partial \varphi(np_n^{1,2})}{\partial \boldsymbol{p}_n^1}\frac{\partial \boldsymbol{p}_n^1}{\partial \boldsymbol{r}^0} = 0.$$

    (iii)    For $\alpha = 1$ and odd n, we have

$$\phi_n^1 = E_n^{el_1} + E_n^{el_2}. \tag{A.15}$$

$$\frac{\partial U}{\partial \boldsymbol{r}^1} = \frac{\partial \phi_n^1}{\partial \boldsymbol{r}^1} = \frac{\partial E^{el_1}}{\partial \boldsymbol{r}^1} + \frac{\partial E^{el_2}}{\partial \boldsymbol{r}^1}. \tag{A.16}$$

$$\frac{\partial U}{\partial \boldsymbol{r}^1} = \frac{\partial \varphi(n r^{1,0})}{\partial \boldsymbol{r}^1} + \frac{\partial \varphi(n r^{1,2})}{\partial \boldsymbol{r}^1} = 0. \tag{A.17}$$

(iv) For $\alpha = 0$ and odd n, we have

$$\phi_n^0 = E_n^{el_1} + E_n^{0,-n} \tag{A.18}$$

Then, considering the Eq. (97), we have $\partial \boldsymbol{p}_n^M / \partial \boldsymbol{r}^0 = 1$ and $\partial \boldsymbol{p}_n^1 / \partial \boldsymbol{r}^0 = 1/n$. Thus

$$\frac{\partial U}{\partial \boldsymbol{r}^0} = \frac{\partial \phi_n^0}{\partial \boldsymbol{r}^0} + \frac{\partial \varphi(n p_n^{M,M-1})}{\partial \boldsymbol{p}_n^M} \frac{\partial \boldsymbol{p}_n^M}{\partial \boldsymbol{r}^0} + \frac{\partial \varphi(n p_n^{1,2})}{\partial \boldsymbol{p}_n^1} \frac{\partial \boldsymbol{p}_n^1}{\partial \boldsymbol{r}^0} \tag{A.19}$$

$$= \frac{\partial \varphi(n r^{1,0})}{\partial \boldsymbol{r}^0} + \frac{\partial \varphi(r^{0,-n})}{\partial \boldsymbol{r}^0} + \frac{\partial \varphi(n p_n^{M,M-1})}{\partial \boldsymbol{p}_n^M} \frac{\partial \boldsymbol{p}_n^M}{\partial \boldsymbol{r}^0} + \frac{\partial \varphi(n p_n^{1,2})}{\partial \boldsymbol{p}_n^1} \frac{\partial \boldsymbol{p}_n^1}{\partial \boldsymbol{r}^0} = 0.$$

*A.3. Energy consistency*

Finally, we show that by the introduced procedure of adding nodes and elements to the multiscale model, the true energy is recovered. Let $E_n$ be the energy between $n$th-nearest neighbors. Then, we investigate the energy consistency for even and odd $n$ separately. For even $n$, the missing energy is the energy not taken into account by the atomistic side, in addition to half of the energy of the element $el_1$ which is also not accounted for due to the second term in Eq. (94). Thus

$$E_n^{miss} = \frac{1}{2}(n-1)E_n + \frac{1}{2}E_n. \tag{A.20}$$

Adding energy by elements according to the pattern in Fig. (29) and noting that the number of elements for the $n$th row is $M - 1$ which is given by Eq. (95) in terms of $n$, the added energy to the system is equal to

$$E_n^{add} = \left(\frac{n+2}{2} - 1\right) E_n, \tag{A.21}$$

and hence the trade-off results in

$$-E_n^{miss} + E_n^{add} = 0. \tag{A.22}$$

For odd $n$, the missing energy is the energy not taken into account by the atomistic side (note that for this case, the energy of the element $el_1$ is fully computed). Thus

$$E_n^{miss} = \frac{1}{2}(n-1)E_n. \tag{A.23}$$

Likewise, considering Eq. (95), the added energy can be written as

$$E_n^{add} = \left(\frac{n+1}{2} - 1\right) E_n, \tag{A.24}$$

and hence the trade-off results in

$$-E_n^{miss} + E_n^{add} = 0 \tag{A.25}$$